\documentclass{article}

\usepackage{arxiv}
\usepackage[T1]{fontenc}    
\usepackage{hyperref}       
\usepackage{url}            
\usepackage{booktabs}       
\usepackage{amsfonts}       
\usepackage{nicefrac}       
\usepackage{microtype}      
\usepackage{graphicx}
\usepackage{natbib}
\usepackage{doi}
\usepackage{newtxtext}
\usepackage{newtxmath}
\hypersetup{
    colorlinks = true,
    urlcolor   = blue,
    citecolor  = black,
}

\newcommand{\RomanNumeralCaps}[1]
\linenumbers

\newcommand{\RE}{\mbox{\textrm Re}}
\newcommand{\ST}{\mbox{\textrm St}}
\newcommand{\RExs}{\RE_{x,s}}
\newcommand{\STREInline}{\ST/\RE^{1/2}}

\newcommand{\STeREInline}{\ST_e/\RE^{1/2}}
\newcommand{\STstareREInline}{\ST^*_e/\RE^{1/2}}
\newcommand{\STstareREDisplay}{\frac{\ST_e^*}{\sqrtREDisplay}}
\newcommand{\STeREDisplay}{\frac{\ST_e}{\RE}}
\newcommand{\sqrtREInline}{\RE^{1/2}}
\newcommand{\sqrtREDisplay}{\sqrt{\RE}}
\newcommand{\sqrtRExsInline}{\RExs^{1/2}}
\newcommand{\sqrtRExsDisplay}{\sqrt{\RExs}}

\newcommand{\sqrtDisplay}[1]{\sqrt{#1}}

\title{Aerodynamic Control of Laminar Separation on a Wall-Bounded Airfoil at Transitional Reynolds Numbers}

\author{\href{https://orcid.org/0009-0004-6032-889X}{
            \includegraphics[scale=0.06]{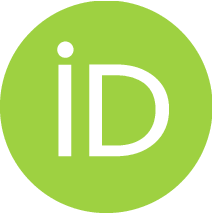}\hspace{1mm}
            Charles Klewicki
            }\thanks{charles.klewicki@usc.edu} \\
        Department of Aerospace and Mechanical Engineering, \\
        University of Southern California, \\
        Los Angeles, CA 90089, USA
	\And
	\href{https://orcid.org/0009-0007-3159-8323}{
            \includegraphics[scale=0.06]{orcid.eps}
            \hspace{1mm}
            Ari Schenkman
            }    \\
        Department of Aerospace and Mechanical Engineering, \\
        University of Southern California, \\
        Los Angeles, CA 90089, USA
	\And
	\href{https://orcid.org/0000-0003-3033-7897}{
            \includegraphics[scale=0.06]{orcid.eps}
            \hspace{1mm}
            Geoffrey R. Spedding
            } \\
        Department of Aerospace and Mechanical Engineering, \\
        University of Southern California, \\
        Los Angeles, CA 90089, USA
}

\renewcommand{\shorttitle}{Aerodynamic Control of Laminar Separation}

\hypersetup{
pdftitle={A template for the arxiv style},
pdfsubject={q-bio.NC, q-bio.QM},
pdfauthor={David S.~Hippocampus, Elias D.~Striatum},
pdfkeywords={First keyword, Second keyword, More},
}

\begin{document}
\maketitle

\begin{abstract}
	Experiments were conducted in a low-turbulence wind tunnel to investigate the efficacy of localised acoustic forcing upon the dynamics and stability of the flow on a cambered, wall-bounded airfoil over a range of Reynolds numbers (Re) where the flow state can switch between two limits -- a low-lift state (SI) where  separation continues beyond the trailing edge and a high-lift state (SII) where the separated flow is closed off to form a laminar separation bubble. The switching between SI and SII can occur close to a critical angle of attack ($\alpha_{\textrm{crit}}$) which varies with $\RE$.  The most effective forcing frequencies are found at a constant value of a rescaled Strouhal number, $\ST^* = \STREInline= 0.027$, which indicates that though the primary unstable modes of the separated shear layer are of the inviscid, Kelvin-Helmholtz type, these modes are seeded by length scales that originate in the laminar (viscous) boundary layer. The most effective chordwise forcing location varies with $\STREInline$ and incidence angle, $\alpha$, and is always upstream of the separation point.  Although the boundary layer flows are far from two-dimensional, forcing at a fixed chord location across all spanwise locations is effective in controlling the SI -- SII transition. Strategies for active and passive feedback control are suggested.
\end{abstract}

\section{Introduction}
\label{sec:intro}

\subsection{Transitional Aerodynamics}

As the chord-based Reynolds number, $\RE = Uc/\nu$ ($U$ is the flight speed, $c$ is the chord length, and $\nu$ is the kinematic viscosity), of a flying vehicle falls below about $10^6$, the aerodynamic properties become strongly influenced not only by the angle of incidence, $\alpha$, but also $\RE$ itself (which is the source of only minor variations when $\RE > 10^6$).  In particular, in the range $\RE = 10^4 - 10^5$, there is a sometimes delicate balance as the laminar boundary has relatively weak resistance to separation but can also reattach, in a time-averaged sense, thanks to the rapid development of shear-layer instabilities and their subsequent transition to turbulence.  There is a strong dependence of performance metrics such as the lift to drag ratio, $L/D$, on whether the flow does or does not reattach; and there is considerable interest in the origin of separation as well as the development of initially small perturbations in the detached shear layer. When the shear layer instabilities are of sufficient amplitude, either in the form of spanwise coherent vortices or as turbulent fluctuations that bring high-speed fluid close to the wall, time-averaged reattachment occurs. The effective airfoil shape then includes the streamline curvature as the separated flow is closed off before the trailing edge, and this closed, recirculating region is termed a laminar separation bubble, or LSB \citep{tani:64}.  As noted by \cite{lissaman:83}, the dynamics of the LSB then dictate the aerodynamic performance of the wing section.

For a uniform, turbulence-free condition, far from boundary effects, the performance of a given wing section is a function of $\RE$ and $\alpha$.  The effects of $\RE$ variation were demonstrated by \cite{zaman:91}, showing abrupt jumps in $C_L$ as $\RE$ rose through the range of $5 - 7 \times 10^4$ for two airfoils (LRN-(1)-1007 and Wortmann FX 63-137), and more recently by \cite{toppings:24} on a NACA 0018 across a range of $3^\circ\leq\alpha\leq 9^\circ$ by varying $\RE$ through $=4.8-8.6 \times 10^4$ in a single time series. Doubling or halving of $C_L$ was observed when $\RE$ was increased or decreased around a critical value where LSB formation would occur.

The variation with $\alpha$ was examined by \cite{tank:21} on a NACA 65(1)412 at $\RE =2 \times 10^4$. A critical angle of attack ($\alpha_\textrm{crit}$) was identified at which point the wing would experience an abrupt, non-linear jump in lift and drop in drag if $\alpha$ were increased by an increment as small as $0.5^\circ$. Through time-averaged PIV fields, \cite{tank:21} connected the discontinuity in the lift and drag polars to a change in flow state from one with laminar separation without reattachment (SI), to one with laminar separation with reattachment (SII), much as previously noted for the Eppler 387 by \cite{yang:14}

In addition to the governing parameters $\RE$ and $\alpha$, small variations in geometry and/or freestream conditions are known to influence the flow state. Both \cite{jaroslawski:2023} and \cite{istvan:2018} found that that higher freestream turbulence intensities promote transition, resulting in the formation and shrinking of LSBs. Such sensitivities likely are responsible for the difficulty in reconciling experimental and numerical results for ostensibly similar conditions \citep{selig:95, yang:13c, tank:17}.

A number of studies have investigated the mechanisms of transition in the shear layer originating from laminar separation \citep{burgmann:08a, jones:10, rodriguez:13, kurelek:21}. The boundary layer separates from the airfoil surface and initially remains laminar, creating a stagnant zone termed the dead-air region by \cite{Horton:68}. As the separation line further diverges from the surface, reverse flow velocities increase and a strong shear layer develops. Exponential growth of disturbances causes the shear layer to destabilize into a two-dimensional instability similar to a Kelvin-Helmholtz (K-H) \citep{kelvin:1871, Helmholtz:1868} instability, albeit with a nearby boundary. \cite{diwan:09} experimentally investigated laminar separation on a flat plate and found that the origin of perturbation growth was due to inflectional instabilities upstream of separation. Studies by \cite{probsting:15a} and \cite{michelis:18}  have also located the amplification of disturbances as occurring upstream of separation and confined to a select frequency range. The inflectional instabilities identified by \cite{diwan:09} were found to be convective in nature and the dominant mechanism for the growth of disturbances into the dead-air region. After detachment of the boundary layer and sufficient distancing of the separation streamline from the airfoil surface, the dominant amplification mechanism of disturbances becomes that of an inviscid shear layer instability, with higher growth rates. Similar findings by \cite{yarusevych:06, jones:08} and \cite{boutilier:12} identified these K-H  shear layer instabilities as the primary mechanism for disturbance growth through experiment and linear stability theory. In contrast to the convective instabilities discussed by \cite{diwan:09}, studies by \cite{alam:00} and \cite{dellacasagrande:24} identified the onset of absolute instabilities for reverse flows of sufficient height or magnitude within the separated region. In either case,exponential growth of disturbances leads to two-dimensional roll-up vortices which initially remain coherent. Eventually, these roll-up vortices are destabilized by nonlinear interactions as they convect downstream and rapidly transition to turbulence. 

Direct Numerical Simulations by \cite{klose:25} detail the three-dimensional transition process for a NACA 65(1)412, noting the appearance and merging of braid vortices (also discussed by \cite{ho:84} in relation to K-H instabilities) between successive roll-up vortices as the beginning of spanwise instabilities breaking up the spanwise coherent rollers. The presence of {C}row instabilities \citep{Crow:70} brought on by trailing edge vortex interactions is also observed in the near-wake, further illustrating the range of physical mechanisms involved with transition over an airfoil. This brief survey gives context for the rich set of dynamics and flow structures present in transitional aerodynamic flows and the role their sensitivities play in the abrupt performance changes experienced by wings in this $\RE$ regime.

\subsection{Tonal Noise Emission and The  Acoustic Feedback Loop}

The propensity of the shear layer to be affected by perturbations at specific frequencies, in conjunction with the unique possibility of tonal noise emission from airfoils at transitional Reynolds numbers, enables a self-sustaining process known as the acoustic feedback loop (AFL) \citep{longhouse:78, arbey:83, golubev:21}. Essential to the acoustic feedback loop is the presence of airfoil tonal noise, most commonly at low $\alpha$, when coherent vortex shedding at the trailing edge is spanwise uniform  and periodic. Studies \citep{desquesnes:07, depando:14, probsting:15b} have investigated tonal noise emissions from airfoils and \cite{golubev:21} presents maps of airfoil noise emission for different Reynolds numbers and angles of attack. \cite{probsting:15a} showed that tonal noise will occur when coherent vortices are shed over the trailing edge of an airfoil from either the suction or pressure surface. When these vortices convect past the trailing edge, they will scatter and produce acoustic waves at the frequency of the vortex shedding. These waves will propagate in all directions, but importantly back upstream, where the most receptive regions are either at the maximum edge velocity or at the onset of instability in the laminar boundary layer, upstream of separation.

 The presence of tonal noise is not required for the formation of an LSB, as pointed out by \cite{jones:08}, and conversely, the formation of an LSB does not guarantee tonal noise emission, as shown by \cite{probsting:15a}. Rather, the acoustic feedback loop will take place if the frequencies of the acoustic waves shed from the trailing edge are in the vicinity of unstable shear layer modes. When the acoustic waves interact with these unstable modes a phase lock between the acoustic wave frequency and shear layer shedding occurs \citep{probsting:15a, thomareis:18}.

\subsection{Acoustic Forcing of Airfoils with Laminar Separation}
The demonstrated sensitivity of the laminar boundary layer to small disturbances, and the known AFL mechanisms encourage controlled forcing methods through acoustic waves, where excitation sources may be external, or internal through miniature speakers embedded in the wing.  A number of studies \citep{zaman:91, yarusevych:06, yang:13a, yang:14, benton:18, celik:23, coskun:24} have identified optimum forcing frequencies, or bands of effective frequencies, and have characterized the effect of varying amplitude and forcing location. Significant improvements in airfoil performance are possible with the implementation of acoustic forcing and marked changes in flow fields occur when excitation frequencies match instability frequencies as shown by \cite{zaman:91} and \cite{yarusevych:07}. 

\subsubsection{Optimal Frequency for Promoting Reattachment}

Numerical and experimental studies have demonstrated that on airfoils with laminar separation, optimal excitation frequencies for promoting transition vary with $\RE$, $\alpha$, and airfoil surface curvature (i.e. the magnitude of the adverse pressure gradient at separation). \cite{zaman:91} first demonstrated that the optimal forcing frequency scaled with $U^{3/2}$  for a given airfoil chord length, and that the most effective frequencies occur when the parameter $\STREInline = 0.02 - 0.03$. The authors also noted that for forcing to remain effective when the flow has switched to the reattached state, then it must impact the boundary layer itself, rather than exclusively a separated layer that may not exist.
\cite{zaman:91} further demonstrated that the range of most effective frequencies shrinks with decreasing $\RE$. Though the $1/\sqrtREInline$ scaling suggests a dependence on laminar boundary layer dimensions, it was noted that significant amplification occurred in the separated shear layer. Based on this and the tonal noise and AFL discussion, an effective forcing causes a phase-lock between the forcing frequency and most amplified modes in the shear layer, via viscous length scales that are initially excited before separation \citep{yarusevych:05, yarusevych:07, coskun:24}.

\subsubsection{Effect of Forcing Amplitude}

Sweeps through varying excitation amplitudes, $A_e$, for known effective excitation frequencies have shown that a relatively small input energy can result in large performance improvements \citep{yarusevych:07, yang:14, michelis:17, coskun:24}. When forced at an excitation frequency, $f_e$, at or close to those that excite the K-H instabilities, $C_L(A_e)$ curves have a sigmoidal shape,  so below a threshold $A_e$, there is only a small effect, and once the flow state has switched, further increases in $A_e$ indicate a saturation in response.

\subsubsection{Optimal Forcing Location}

The most effective acoustic forcing location varies with $\RE$ and separation location $x_s$. \cite{yeh:20} numerically investigated optimal actuator location for harmonic momentum-based forcing on a NACA0012 at $\mathrm{Re}= 2\times10^5$ via resolvent analysis. The unforced flow included a separation bubble spanning $x/c = 0.022-0.08$. It was found that the optimal forcing location 
\[
x^* = \underset{x}{\text{argmax}} \,(|\hat{f}|),
\]
(where $|\hat{f}(x)|$ is the forced response mode magnitude) always occurred upstream of separation for forcing at a variety of frequencies. This finding is supported by the notion that perturbation growth begins upstream of separation, and is in agreement with previous numerical and experimental studies involving forcing of flows with laminar separation \citep{huang:88, hsiao:90, greenblatt:00, yeh:19}. Additionally, \cite{yeh:20} explored optimal orientation of forcing, finding that a surface tangent direction was optimal for momentum flux forcing. At a higher Reynolds number of $\RE=3 \times 10^5$, experiments on a NACA $63_3-018$ airfoil by \cite{hsiao:97} demonstrated delayed stall with leading edge forcing and showed most effective control when the shear-layer shedding ``locked in'' with acoustic forcing and when forcing occurred at the separation point. \cite{yang:14} showed that improvements in $\Delta L/D$ for an Eppler 387 at $\RE = 6\times 10^4$ were independent of actuator location, but the range of frequencies which could yield an increase in $L/D$ shrank as the actuator location moved from $x/c = 0.1$ to $0.6$.

\subsection{Scope and Objectives}

The susceptibility of transitional flows to specific perturbation, combined with the resulting improvements in wing performance upon the promotion of transition and reattachment, renders them suitable, in principle, for control. The aim of this investigation is to expand on the current literature by experimentally investigating chordwise actuator location ($x_e$) on a finite wall-bounded wing across a parameter range of $\RE$ and $\alpha$ as well as excitation frequency ($f_e$) and amplitude ($A_e$). This will be accomplished through direct force measurements and flow field measurements on a NACA 65(1)412 airfoil in the Dryden Wind Tunnel (DWT). Comparisons with numerical and experimental results will be made when possible with the objective of identifying consistent flow structures and optimal actuator placement. 

\begin{figure}
    \begin{center}
    \includegraphics[width=1\textwidth,]{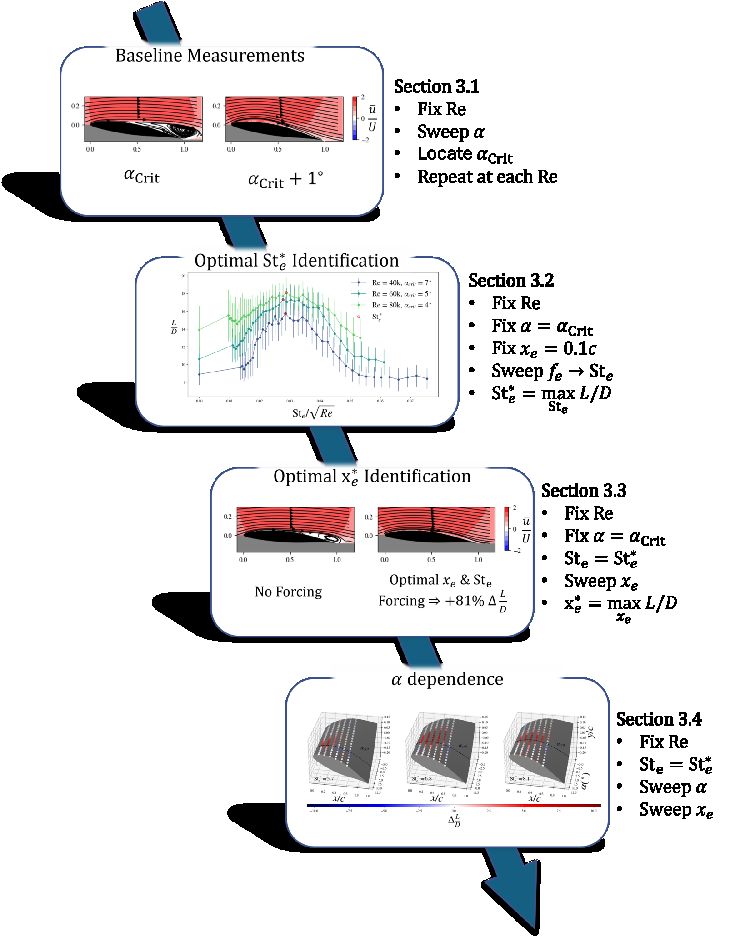}
    \end{center}
    \caption{Experimental procedure for systematic investigation of parameter space.}
    \label{fig:Roadmap}
\end{figure}

Figure \ref{fig:Roadmap} shows a road map for addressing the scope of this study. First, in section \ref{subsec:Baseline Measurements} characterization of the baseline flow is achieved through sweeps of $\alpha$ at $\RE = 2,4,6,8\times 10^4$.  $\alpha_\text{crit}$ is identified for each $\RE$, where the flow switches state between separation without time-averaged reattachment and an LSB. This is followed by forced experiments starting in section \ref{subsec:Forcing Frequency Identification}. Initial frequency sweeps at the forward-most forcing location on the airfoil are conducted for each $\RE$ at $\alpha_\text{crit}$ and optimal forcing frequencies ($\ST^*_e$) are observed, with respect to improvement in airfoil performance ($L/D$). In the subsequent section (\ref{subsec:Forcing Location Trends}) $\ST^*_e$ optimality is demonstrated across streamwise forcing location ($x_e$) and angle of attack ($\alpha$). In addition, section \ref{subsec:Forcing Location Trends} assesses forcing effectiveness as a function $x_e$ and identifies optimal forcing locations $x_e^*$. In section \ref{subsec:Sensitivity to alpha} the $\alpha$ dependence of optimal parameters St$^*_e$ and $x_e^*$ are investigated by sweeping through $x_e$ and $\alpha$ at each $\RE$ while fixing the forcing frequency at St$^*_e$. Finally, follow-up experiments are presented in section \ref{subsec:39 speakers}. These results---and their implications and limitations---are discussed in section \ref{sec:discussion}.

\section{Experimental Methods}
\label{sec:methods}

\begin{figure}
    \centering
    \includegraphics[width=1\linewidth]{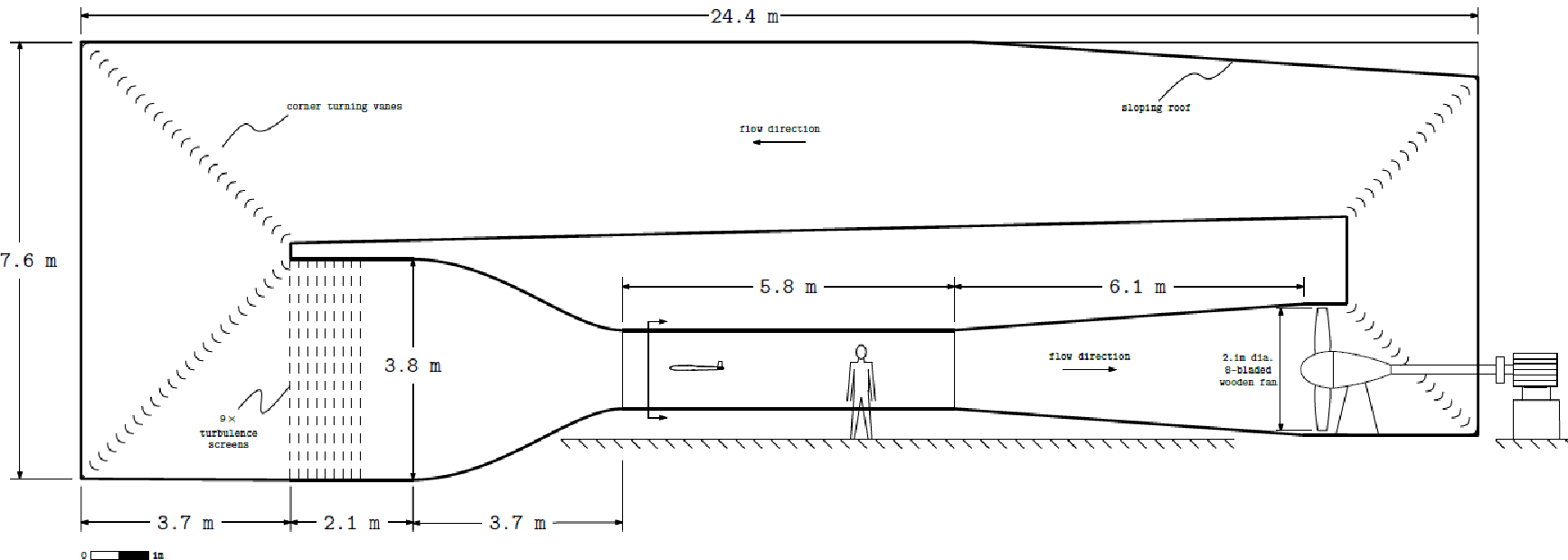}
    \caption{Dryden Wind Tunnel Schematic.}
    \label{fig:DWT Schematic}
\end{figure}

Experiments were conducted in the Dryden Wind Tunnel (DWT). The DWT is a recirculating atmospheric wind tunnel with three octagonal test sections in series. The wall to wall width of each test section is 1.37 m and a schematic of the wind tunnel side-view is shown in figure \ref{fig:DWT Schematic}. The wind tunnel has 9 mesh screens to break down turbulence before entering the settling chamber, with a 7:1 contraction ratio from the settling chamber to the first test section. The turbulence intensity has been measured by \cite{tank:18} to be $T  \leq 0.035\%$ for $10 \leq f \leq 200$ Hz in the velocity range $4 \leq U \leq 20$ m/s.

\subsection{Airfoil Model}

\begin{figure}
    \centering
    \includegraphics[width=0.9\linewidth]{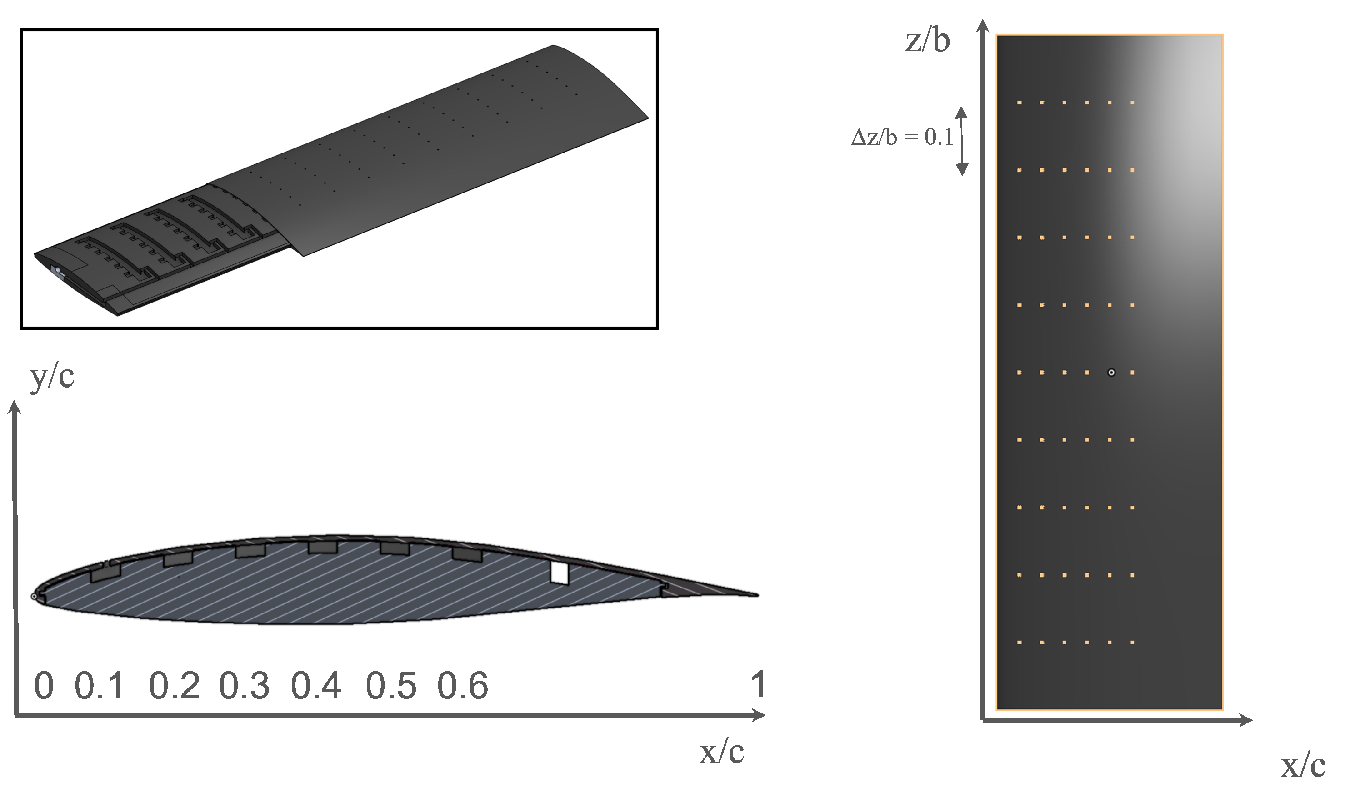}
    \caption{Two-piece NACA 65(1)412 model. (Top Left): Isometric view of model with suction surface partially mounted with inside of bottom carriage visible. (Bottom Left): Cross section view with speaker cavities and wire channel visible. (Right): Top view of suction surface with speaker holes highlighted.}
    \label{fig:Two piece NACA 65(1)412 model}
\end{figure}

A NACA 65(1)412 profile was used to generate two models, both with a chord length  $c = 200$ mm and, span $b = 600$ mm. The first model was a one-piece PLA construction and the second was a two-piece PLA model created for internal acoustic excitation. The bottom pressure side carriage of the two-piece model contained channels and cavities in which speakers and associated wiring could be placed (Figure \ref{fig:Two piece NACA 65(1)412 model}). The top piece was a thin shell suction side that slid onto the carriage and could be easily modified and replaced. Each model was sanded to a P1000 grit finish, alternately sanding and painting with a sandable filling primer to ensure a smooth surface.   

\subsection{Experimental Configuration}

\begin{figure}
    \begin{center}
    \includegraphics[width=0.75\textwidth,]{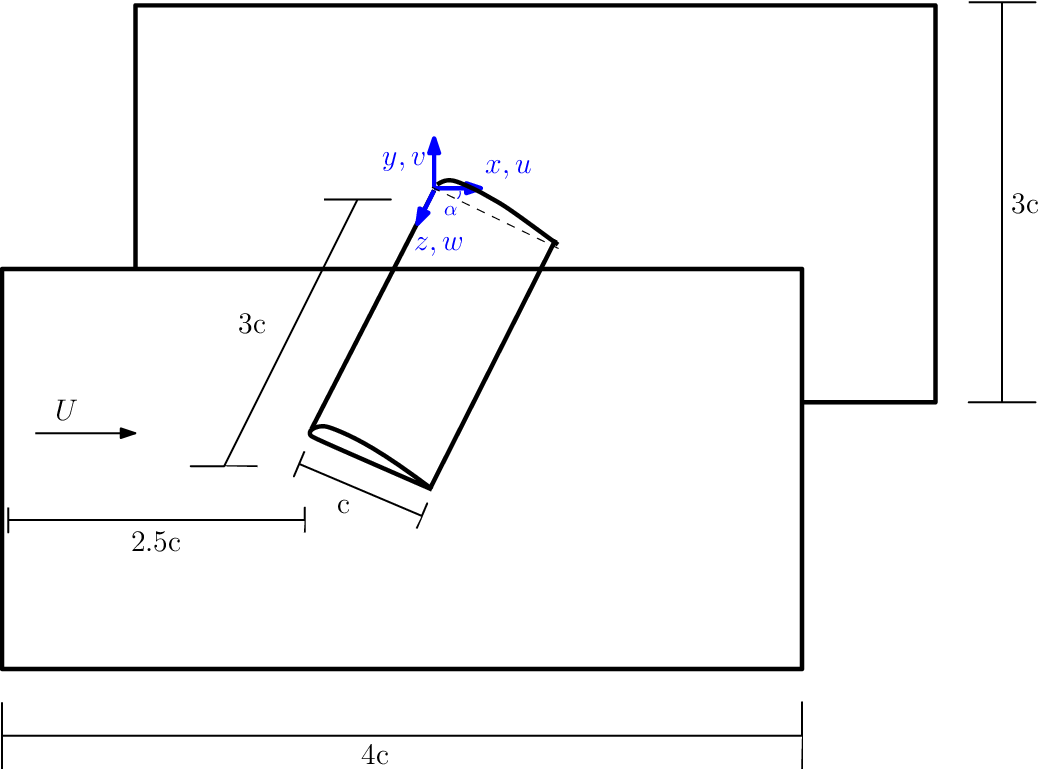}
    \end{center}
    \caption{Model-based coordinate system and geometry of wall-bounded airfoil.}
    \label{fig:coordinate system}
\end{figure}

The models were mounted vertically between two end walls with dimensions 1200 x 600 x 4.8 mm$^3$. The top end wall was transparent to allow for PIV and flow visualizations. The bottom end wall was black to reduce light reflections in the wind tunnel. The leading and trailing edges of the end walls were tapered to a right--triangular edge with an angle of $60^\circ$. The coordinate system had its origin at the leading edge of the airfoil with $x$ and $y$ axes parallel and perpendicular to the freestream direction and the $z$ axis traversing the span from end wall to end wall (Figure \ref{fig:coordinate system}). 

\begin{figure}
    \centering
    \includegraphics[width=1\linewidth]{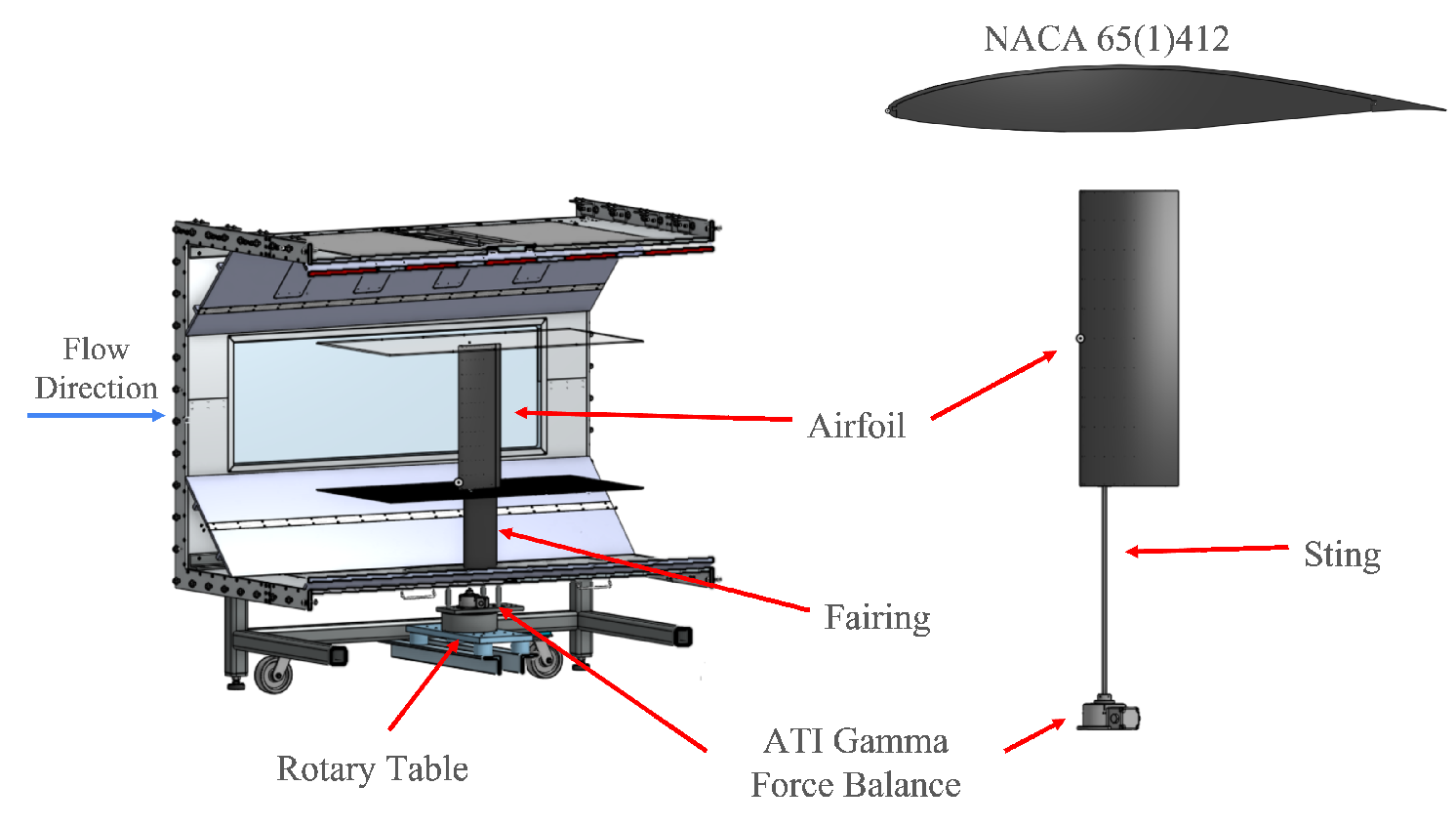}
    \caption{The NACA 65(1)412 model and force balance configuration.}
    \label{fig:CAD Wind Tunnel}
\end{figure}

The wing was supported by a sting which was mounted to an ATI Gamma force balance for direct measurements of aerodynamic forces. A shroud was placed around the section of the sting exposed to the freestream in order to isolate the forces experienced by the wing. The force balance was mounted to a Parker PM-DDD12DNOH rotary table allowing for $\alpha$ sweeps. The assembled experimental setup is shown in figure \ref{fig:CAD Wind Tunnel}.  

The airfoil $\alpha$ was aligned with the freestream by setting up a calibration plane parallel with the freestream using a Bosch GLL 55 self-leveling cross-line laser. A PLA negative contour of the airfoil was mounted near mid-span on the airfoil such that the mounting screws intersected the laser sheet. Using a camera to locate the intersection of the laser plane and the screws, the airfoil was rotated until the laser intersected both screws at the same location. This location was considered $\alpha = 0^\circ$. As the uncertainty in laser plane--screw intersection was on the order of 0.1 mm, the uncertainty in the $\alpha=0^\circ$ designation is on the order of $0.03^\circ$

The mean reference speed, $U$, was measured with a pitot tube, mounted 1 m upstream of the wing, and connected to a 0.5INCH-D2-4V-MINI differential pressure sensor.

\subsection{Force Balance}

An ATI Gamma force balance with an SI-32-2.5 calibration was used to capture the forces and torques on the airfoil. The range and resolution of the balance are given in table \ref{table:Gamma Cal} where subscripts $A$ and $N$ represent the axial and normal directions based on the chord aligned coordinate system. Prior to experimentation, estimates of lower limiting forces on the wing were made from previous data \citep{tank:18}, and estimated to be 5 mN for $\RE=2\times10^4$ and 2.9 N for $\RE=8\times10^4$. The minimum force estimate is just below the resolution of the force balance resulting in a signal to noise ratio close to 1 for drag data near $\alpha = 0^\circ$ at $\RE=2\times10^4$. To mitigate the limitations of the force balance at such low speeds, statistics were compiled through repeated experiments using different mountings and models. The results were averaged and uncertainties were reported as test-to-test variation about the mean.

\begin{table}
    \centering
    \renewcommand{\arraystretch}{2}
    \begin{tabular}{cccc|cccc}
         &Range&&&&Resolution&& \\
         $F_{A,N}$ &$F_z$&$\tau_{A,N}$&$\tau_z$&$F_{A,N}$ &$F_z$&$\tau_{A,N}$&$\tau_z$ \\
         \hline
         32 N & 100 N & 2.5 N-m & 2.5 N-m & 6.25 mN & 12.5 mN & 5 mN-m & 5 mN-m
    \end{tabular}
    \caption{Resolution and range of ATI Gamma Force / Torque sensor.}
    \label{table:Gamma Cal}
\end{table}

The measured normal and axial forces were rotated to conventional lift and drag components by
\begin{align*}
    L &= F_N \cos(\alpha - \gamma) - F_A \sin(\alpha - \gamma), \\
    D &= F_A \cos(\alpha - \gamma) + F_N \sin(\alpha - \gamma),
\end{align*}
where $L$ is lift, $D$ is drag and $\gamma$ is the alignment offset.
Weight tares were conducted before the experiments to remove the components of the force balance signal due to the weight of the wing. These were applied before the forces were transformed.

To compensate for any misalignment between the force balance axes and airfoil axes, a pure drag force calibration was applied to the strut after mounting and aligning $\alpha$. The force was applied by attaching a string to the strut with the other end of the string hung over a pulley and attached to a weight. The string and pulley were aligned to the center line of the wind tunnel with the Bosch GLL 55 for a pure drag force. Force measurements were taken at one degree increments for a hysteresis loop from $-5^\circ \leq \alpha \leq 5^\circ$. The difference between the angle measured by the force balance and $\alpha_0$ was computed using 
\begin{gather}
    \gamma = \alpha_0 - \tan\left(\frac{F_N}{F_A}\right),
\end{gather}
and the value was averaged for the four test points nominally at $\alpha=0^\circ$. 

Gaps between the airfoil and end walls were necessary to isolate the force measurements on the airfoil. These gaps were  measured at the leading and trailing edge for each experiment. Typical end wall gaps were $\leq 3$ mm ($\leq 0.005b$) in accordance with \cite{barlow:99}. Yaw estimates were computed from the end wall measurements and were $\leq0.5^\circ$ for all experiments. Roll estimates were made by aligning the Bosch GLL 55 with the trailing edge at $z=b$ and measuring its deviation from vertical at $z=0$. The roll was $\leq1.4^\circ$ for all experiments.

Force measurements were taken with a sampling frequency of $f=2000$ Hz, equating to $\Delta tU/c \leq 0.015$. $\alpha$ sweeps at fixed Reynolds numbers were conducted by setting the calculated wind speed for a given $\RE$ based on measured temperature, pressure, and relative humidity, waiting 120 s for the wind tunnel to reach a steady state, and then acquiring data. Data were acquired for 10 s ($\geq 75 c/U$) at each $\alpha$ with 20 s ($\geq 150 c/U $) settling times for the flow and mechanical setup to reach steady state after moving to a new $\alpha$; for runs where $\alpha$ did not vary, such as $\ST_e$ sweeps, a 5 s ($\geq 35 c/U $) settling time was used instead. $\alpha$ sweeps were conducted as hysteresis loops with $\alpha$ starting at $\alpha = 0^\circ$ incrementing up to $\alpha_{\max}$, then down to $\alpha_{\min}$, and finally back to $\alpha = 0^\circ$. 

The force measurements were normalized by the dynamic pressure $q$ and the wing planform area $S = bc$:
\begin{align*}
    C_L &= \frac{L}{qS}, \\
    C_D &= \frac{D}{qS}.
\end{align*}

\subsection{Acoustic Forcing and Measurements}
\label{Methods:Acoustic Forcing and Measurements}

Internal acoustic forcing experiments were conducted with speakers mounted inside the wing at a single chordwise location and nine uniformly spaced spanwise locations. The excitation voltage, $V(t)$, was
\begin{gather}
    V(t) = A_e\sin(2\pi f_e t),
\end{gather}
\begin{figure}
    \centering
    \includegraphics[width=0.75\linewidth]{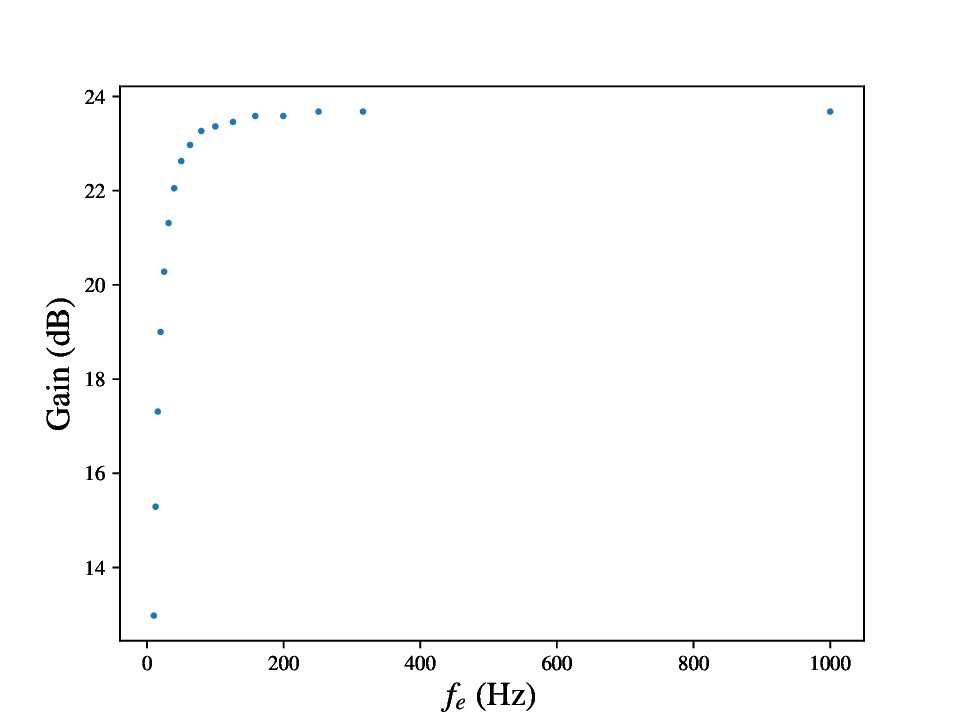}
    \caption{Frequency response of UA-R3010 internal excitation amplifier}
    \label{fig:UA-R3010 Amp Response}
\end{figure}
where $f_e$ is the excitation frequency and $A_e$ is the amplitude. During experiments at a given chordwise location, the remaining unused cavities were filled with PLA printed blocks to prevent acoustic resonances within the empty cavities. The unused speaker holes on the suction surface were filled with putty and lightly sanded to minimize their surface roughness impact. Figure \ref{fig:Two piece NACA 65(1)412 model} shows the different chordwise ($x_e/c= 0.1-0.6, \; \Delta x_e/c = 0.1$) and spanwise ($z_e/b = 0.1-0.9, \; \Delta z_e/b = 0.1$) excitation locations. Certain follow-up experiments were conducted with a higher spanwise density of speaker locations ($z_e/b = 0.025-0.975, \; \Delta z_e/b = 0.025$) at the foremost chordwise position, $x_e/c= 0.1$. The speakers were UT-P2020 MEMS speakers with a nominal frequency range from 20 Hz to 20 kHz. The speakers were wired in parallel to ensure the signals were in phase and the speakers were driven by a UA-R3010 amplifier and 33210A Agilent function generator. The wiring for the speakers was routed out of the wing at the bottom end wall and fixed to the sting so the entire setup moved as one rigid body. 

The UA-R3010 is a linear amplifier for piezoelectric speakers with a 23.5 dB gain for a frequency range from 20 Hz-80 kHz. Its nominal 23.5 dB gain was found to drop off significantly at the lower frequencies ($f_e \leq 200$), as shown in figure \ref{fig:UA-R3010 Amp Response}. A lookup table was created with the experimental frequency response data, and to compensate for the drop off, higher input voltages were used for lower frequencies to ensure the speakers always received a signal of $A_e=28.1$ V peak-peak (V$_{PP}$) $\pm0.1$ V.

The sound pressure level (SPL) of the speakers at $x_e/c=0.3$ at an input voltage of $A=28.1$ V$_{PP}$ was measured with a 4954-B 1/4” B\&K free field microphone mounted to a traverse. The test was conducted with the microphone mounted normal to the speaker hole location at a distance of $\leq 0.1$ mm to the suction surface. The microphone was traversed through different spanwise locations. At each location, the speaker was swept through 28 logarithmically spaced points from 20 Hz to 500 Hz, and a Goertzel filter was applied to identify the measured SPL level of the excitation. The results are shown in figure \ref{fig:internal frequency response SPL}.  

Force measurements for a given chordwise actuator location began with an unforced $\alpha$ sweep at the same $\RE$ to check the effect of speaker holes themselves, and to verify agreement with previous unforced, baseline data. Next, $\alpha$ was set to $\alpha_{\textrm{crit}}$ for the test $\RE$ and a discrete frequency sweep from $2.7 \leq$ $\ST_e \leq 15$ was conducted, where $\ST_e = f_ec/U$ is the excitation Strouhal number. Finally, a forced $\alpha$ sweep with $\ST_e=\ST_e^*$ was carried out, where $\ST_e^*$ is the optimal forcing frequency at the given $\RE$. The process was repeated for $\RE=2,4,6,8 \times 10^4$ and then the location of excitation, $x_e$, was changed. $x_e$ was varied from $x/c = 0.1 - 0.6$. $\ST_e^*$ was fixed from the frequency sweep conducted at $x/c = 0.1$. The optimal, or most effective, excitation frequency was defined as $\ST_e$ at which the maximum $L/D$ occurred.

\subsection{Particle Image Velocimetry}
\label{Methods:PIV}

\begin{figure}
    \begin{center}
    \includegraphics[width=1\textwidth,]{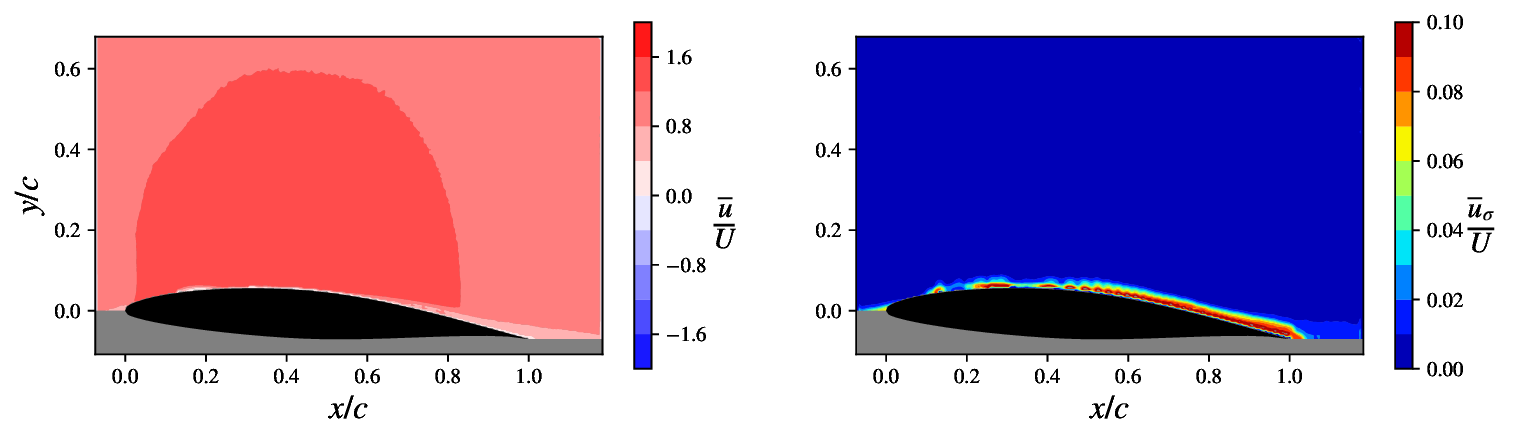}
    \end{center}
    \caption{Representative time averaged velocity field (left) and its corresponding uncertainty distribution (right) for DWT PIV.}
    \label{fig:uncertainty field wind tunnel}
\end{figure}

PIV images were acquired by seeding the wind tunnel with smoke, illuminating the smoke particles with a laser, and acquiring successive image pairs of the particles with a camera. The smoke was injected into the flow in the diffuser of the wind tunnel with a Colt 4 smoke machine. The smoke had a typical particle size of $0.2-0.3$ microns resulting in a settling velocity $\approx8$ mm/hr. A Quantel EverGreen double-pulsed Nd:YAG laser was used in conjunction with two plano-concave lenses in series to produce a laser sheet in the x-y plane and illuminate the particles. The x-y plane was centered above the airfoil suction surface. The laser could produce two 200 mJ pulses with wavelengths of 532 nm at a 15 Hz rate with time between pulses down to 1 $\mu$s. An Imager sCMOS camera with a spatial resolution of 2560 x 2160 pixels and a maximum frame rate of 50 FPS was used to acquire the images. A 50 mm Nikon lens focused at $\approx 0.8$ m with an aperture setting of $f = 2.8$ was mounted on the camera and captured a field of view from the airfoil leading edge to the trailing edge. LaVision’s DaVis software was used to sync the laser and camera for data acquisition, process the images with PIV algorithms to produce instantaneous vector fields, and produce uncertainty estimates. Figure \ref{fig:uncertainty field wind tunnel} shows a representative instantaneous uncertainty field for the PIV measurements. An initial correlation box size of 48x48 pixels with 50\% overlap was reduced to a final correlation box size of 16x16 pixels, with a multi pass FFT window deformation PIV algorithm. This produced vector fields with spatial resolution 0.79 mm ($0.0039c$) in the x and y directions. 

\begin{figure}
    \begin{center}
    \includegraphics[width=1\textwidth,]{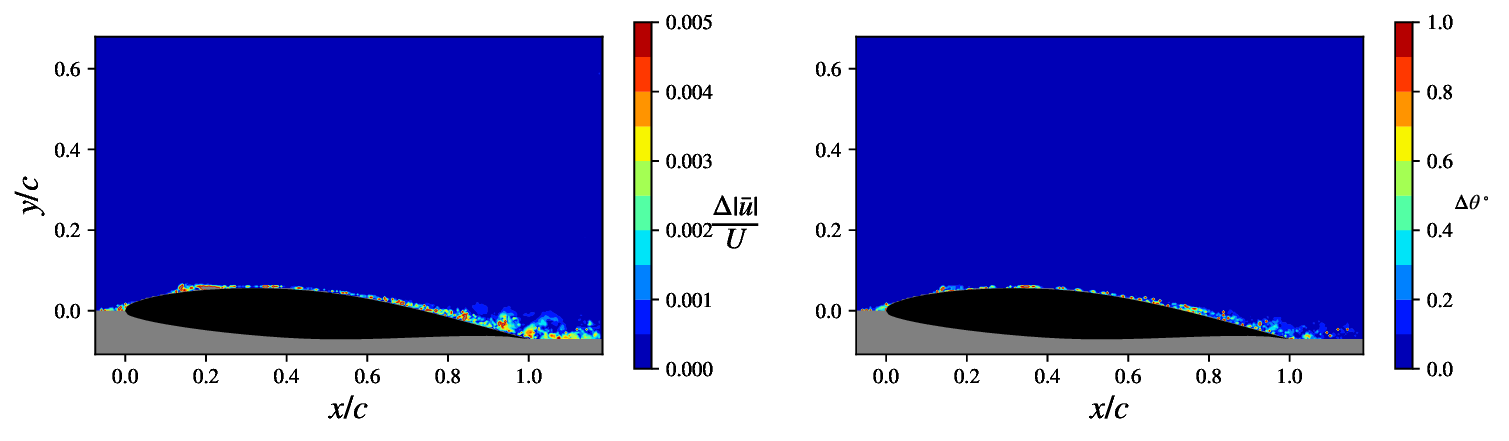}
    \end{center}
    \caption{Difference in vector magnitude (left) and  orientation (right) for time averaged PIV fields with 150 images vs 149 images.}
    \label{fig:PIV convergence}
\end{figure}

The maximum temporal rate of data capture was limited by the laser at 15 Hz making the results non-time resolved since $U\Delta t/c\geq 0.5$ and a convergence study was conducted to determine the number of image pairs necessary for the flow to produce statistically stationary behavior. Two criteria were used: the average vector magnitudes were required to change by max$(\Delta|\vec{u}|/U) < 0.1$ and the vector orientations were required to change by max$(\Delta \theta) <1^\circ$ for the addition of one instantaneous vector field to the statistical population. 150 images were determined to be sufficient. Figure \ref{fig:PIV convergence} shows the difference in magnitude and orientation of the time averaged flow for populations of 149 and 150 images. The non-time resolved nature of the PIV requires the use of ensemble averages which are denoted by an over-bar (e.g. $\bar{u}$)

PIV measurements for the one--piece model were collected during the force measurements at $\alpha =0 ^\circ, 4^\circ, \alpha_{\textrm{crit}}$, and $\alpha_{\textrm{crit}} + 1^\circ$, for $\alpha$ sweeps at all $\RE$. PIV measurements for the two-piece model with internal excitation were captured at $\alpha_{\textrm{crit}}$ for both the forced and unforced cases. Force measurements were taken in tandem to ensure agreement with previous data. Flow field data were taken at mid-span ($z/b = 0.5$).


To relate acoustic excitation location ($x_e$) to boundary layer parameters, estimates of the separation location were computed. The raw PIV spatial resolution was insufficient to accurately identify the separation location as given by the Prandtl separation criterion. Hence, vorticity contours of $\bar{\omega}_zc/U$ were examined and the shear layer was identified at each $x/c$ location downstream of separation by 
\begin{gather}
y_{s} = \underset{y}{\mathrm{argmax}}{|\overline \omega_z |}.
\end{gather}
A cubic polynomial was fit to the points in the shear layer and then extrapolated to the airfoil surface for separation point estimates. Figure \ref{fig:Separation Point Estimates} shows the $\bar{\omega}_z$ contours with the cubic polynomial and separation point estimate. 

\section{Results}
\label{sec:results}

\subsection{Baseline Measurements}
\label{subsec:Baseline Measurements}

\begin{figure}
    \centering
    \includegraphics[width=1\linewidth]{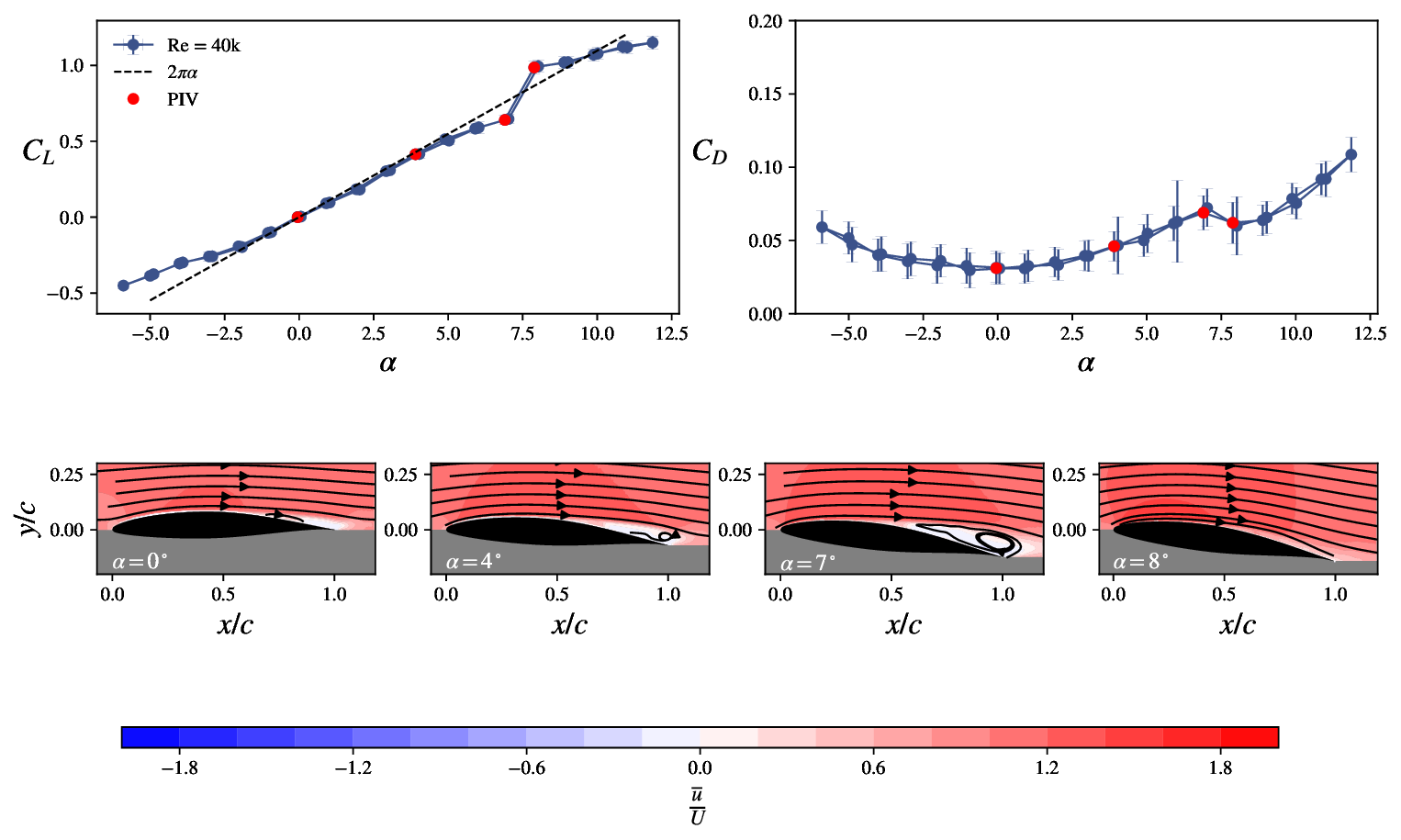}
    \caption{(Top) Unforced baseline lift and drag plots for one--piece PLA model $\RE = 4 \times 10^4$. Red circles denote values corresponding to PIV fields. Uncertainties are the standard deviation of temporal fluctuations about the mean. (Bottom) Corresponding ensemble averaged PIV $\bar{u}$ velocity fields with streamlines for varying $\alpha$.}
    \label{fig:One Piece, Re = 40k, FB+PIV}
\end{figure}

Flow fields around the one--piece airfoil at $\RE = 4 \times 10^4$ are shown in figure \ref{fig:One Piece, Re = 40k, FB+PIV} to illustrate the abrupt transitions in flow state, together with consequences in the time-averaged lift and drag forces. Starting at $\alpha = 0^\circ$ and increasing $\alpha$, $C_L$ linearly increases until $\alpha = 5^\circ$. The respective PIV contours show a separation region near the trailing edge which grows and moves towards the leading edge as $\alpha$ increases. After $\alpha = 5^\circ$, the $C_L$ curve begins to diverge from the $2\pi\alpha$ line with the lift slope, $C_{L\alpha}$, degrading. At $\alpha = 7^\circ$, the separated region reaches its maximum height, and extends well beyond the trailing edge. The lower curvature of the streamlines is associated with the reduction in $C_{L\alpha}$. $\alpha = 7^\circ$ is termed $\alpha_\text{crit}$ \citep{tank:21} denoting that the flow experiences a state change when $\alpha$ is increased beyond $\alpha_\text{crit}$. At $\alpha = 8^\circ$, $C_L(\alpha)$ has increased abruptly, accompanied by a drop in $C_D$. $C_L$ at $\alpha = 8^\circ$ outperforms two-dimensional airfoil theory because the streamlines experience an increase in curvature over the suction surface, as they round the closed separation bubble, increasing circulation, as would occur with a more highly--cambered profile shape. Post-$\alpha = 8^\circ$, both $C_L$ and $C_D$ increase with $\alpha$, but on different curves than the pre-$\alpha_\text{crit}$ state. The performance change in the force measurements after $\alpha_\text{crit}$ is connected to the collapse of the separated region and development of an LSB in the PIV fields.

\begin{figure}
    \centering
    \includegraphics[width=1\linewidth]{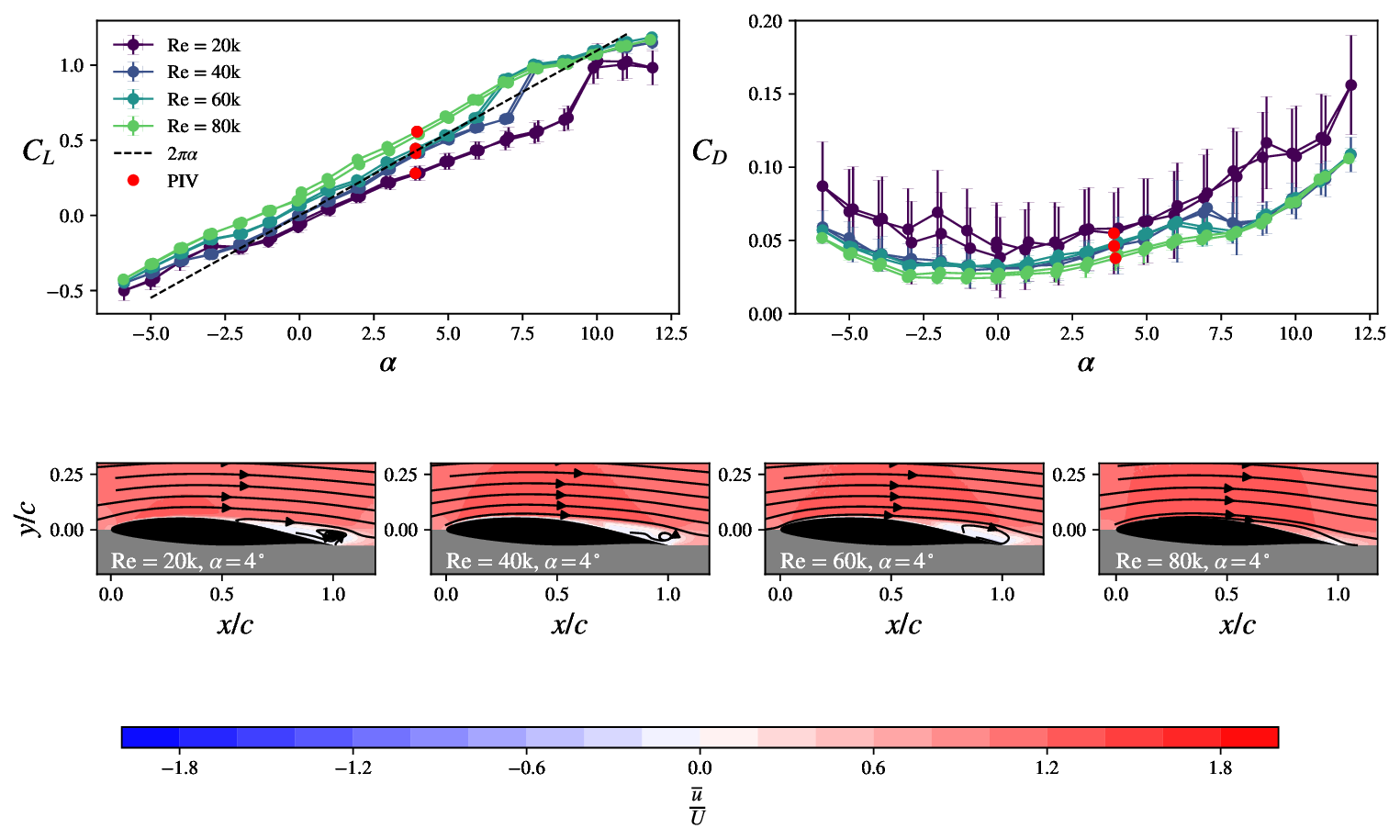}
    \caption{(Top) Unforced baseline lift and drag plots for one--piece PLA model $\RE = 2-8\times 10^4$. Red circles denote values corresponding to PIV fields. Uncertainties are the standard deviation of temporal fluctuations about the mean. (Bottom) Corresponding ensemble averaged PIV $\overline u$ velocity fields with streamlines for fixed $\alpha$.}
    \label{fig:One Piece, Re = 20-80k, FB+PIV}
\end{figure}

Figure \ref{fig:One Piece, Re = 20-80k, FB+PIV} shows the nonlinearities in $C_L(\alpha)$ gradually straightening out as $\RE$ increases.  The $C_L(\alpha)$ curves shift upwards (so at any fixed $\alpha$, $C_L$ increases), and the increase in lift slope, $C_{L\alpha}$ with $\RE$, and decrease in $\alpha_{0L}$ (zero--lift angle of attack) are as noted in \cite{tank:21}.  With the reduced influence of viscosity, $C_D$ at any given $\alpha$ also drops too. The PIV contours demonstrate the thinning of the boundary layer for a fixed $\alpha$ as $\RE$ is increased. Notably, for $\RE= 8\times 10^4$ the separated region is very thin in the PIV contour and does not push the curving of the streamlines into the airfoil wake. This is likely true for all $\alpha$ at $\RE= 8\times 10^4$ and explains the lack of distinct $\alpha_\text{crit}$. At $\RE= 2\times 10^4$ the $C_L$ curve begins to decrease after the switch in flow state and diverges from the collapsed results of the higher $\RE$ cases. Previously, it has been shown \citep{klewicki:25} there is an onset of turbulent separation downstream of the LSB which explains the decrease in $C_L$.

\begin{figure}
    \centering
    \includegraphics[width=1\linewidth]{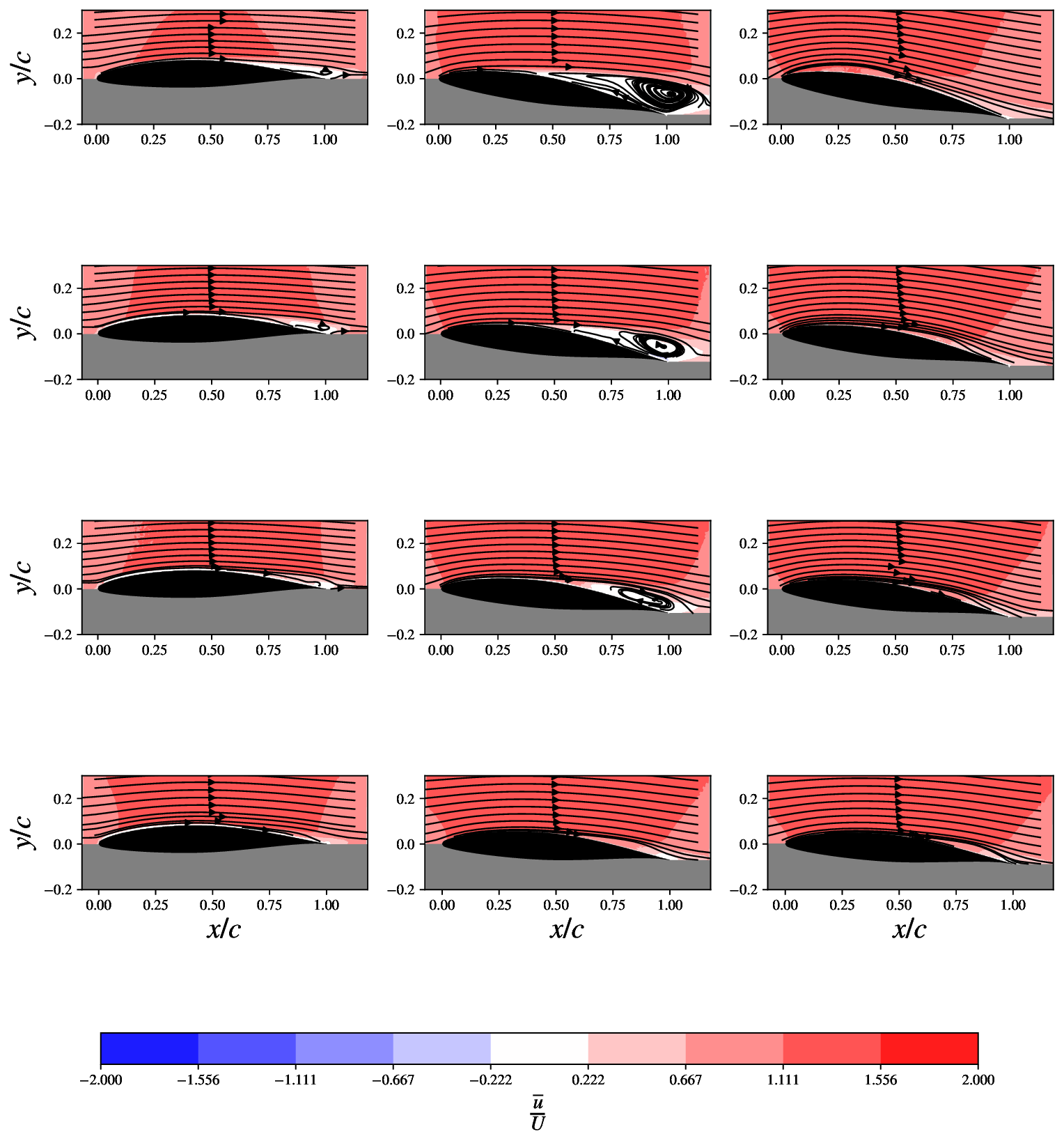}
    \caption{Ensemble averaged PIV $\overline u$ velocity fields with streamlines for  $\alpha = 0, \alpha_\text{crit}, \alpha_\text{crit}+1^\circ$ (Left to Right) and $\RE = 2,4,6,8 \times 10^4$ (Top to Bottom).}
    \label{fig:One Piece, Re = 20-80k, PIV}
\end{figure}

The contours and streamlines in figure \ref{fig:One Piece, Re = 20-80k, PIV} at $\alpha = 0^\circ$ demonstrate the boundary layer thinning as $\RE$ is increased (top to bottom) with a shrinking separated region near the trailing edge. At $\alpha = \alpha_\text{crit}$ this becomes more apparent as the separation regions reach their maximum extent for each $\RE$ and finally at $\alpha = \alpha_\text{crit} +1^\circ$ the switch from open separation to the formation of a closed LSB is observed. The streamlines for the $\RE = 2\times 10^4$ case (top row) show a separation bubble which has formed on the first half of the chord. This is distinct from the results at higher $\RE$ which show LSBs initially forming closer to the trailing edge on the downstream half of the chord. The lower $\RE= 2\times 10^4$ increases the stability of the separated shear layer which in turn requires more development in the streamwise direction for vortex roll up to occur, leading to a higher $\alpha_\text{crit}$. This trend should hold in general regardless of airfoil contour. For this particular airfoil, the maximum thickness at $x/c \approx 0.4$ allows for a relatively large leading edge LSB at $\RE= 2\times 10^4$, while airfoils with maximum thicknesses closer to the leading edge may lead to a more conventional bifurcation of large LSBs on the aft part of the chord and small leading edge LSBs. 

\begin{figure}
    \centering
    \includegraphics[width=1\linewidth]{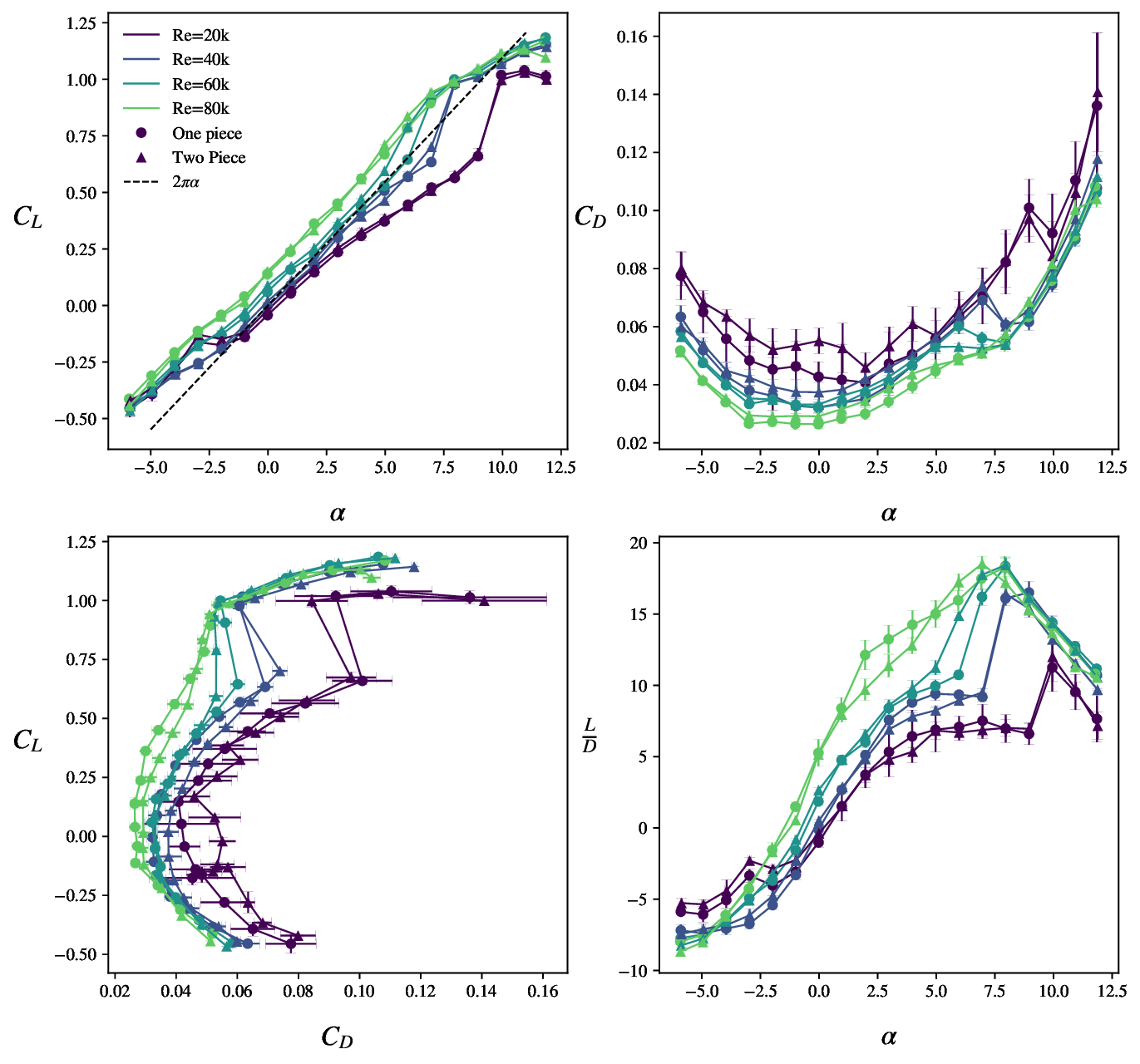}
    \caption{Unforced baseline lift and drag data for one--piece ($\bullet$) and two-piece ($\blacktriangle$) models. The two-piece model was configured with no speaker holes in the suction surface. Uncertainties are the standard deviation of test-test variation of the mean. The theoretical $2\pi\alpha$ line intersects the origin. }
    \label{fig:Drag Polars 1vs2 piece}
\end{figure}

To check the possible influence of small geometric disparities in different fabrication strategies, Figure \ref{fig:Drag Polars 1vs2 piece} compares the force measurements of the one- and two-piece airfoil models. The principal differences lie in $C_D$, which is different beyond experimental uncertainty at low $\RE$ and $\alpha$, and also different for the highest $\RE$ and $\alpha$ below $\alpha_{\text{crit}}$. Given that this flow is known to be highly sensitive to small variations in geometry, the relatively small differences observed here are likely due to slightly higher effective surface roughness on the two-piece construction. Otherwise, the small influence allows us to continue while using the two-piece airfoil as a baseline case for remaining experiments and the $\alpha_{\text{crit}}$ for each $\RE$ will be identified based on the two-piece wing data in figure \ref{fig:Optimal forcing frequencies}.

\begin{figure}
    \centering
    \includegraphics[width=1\linewidth]{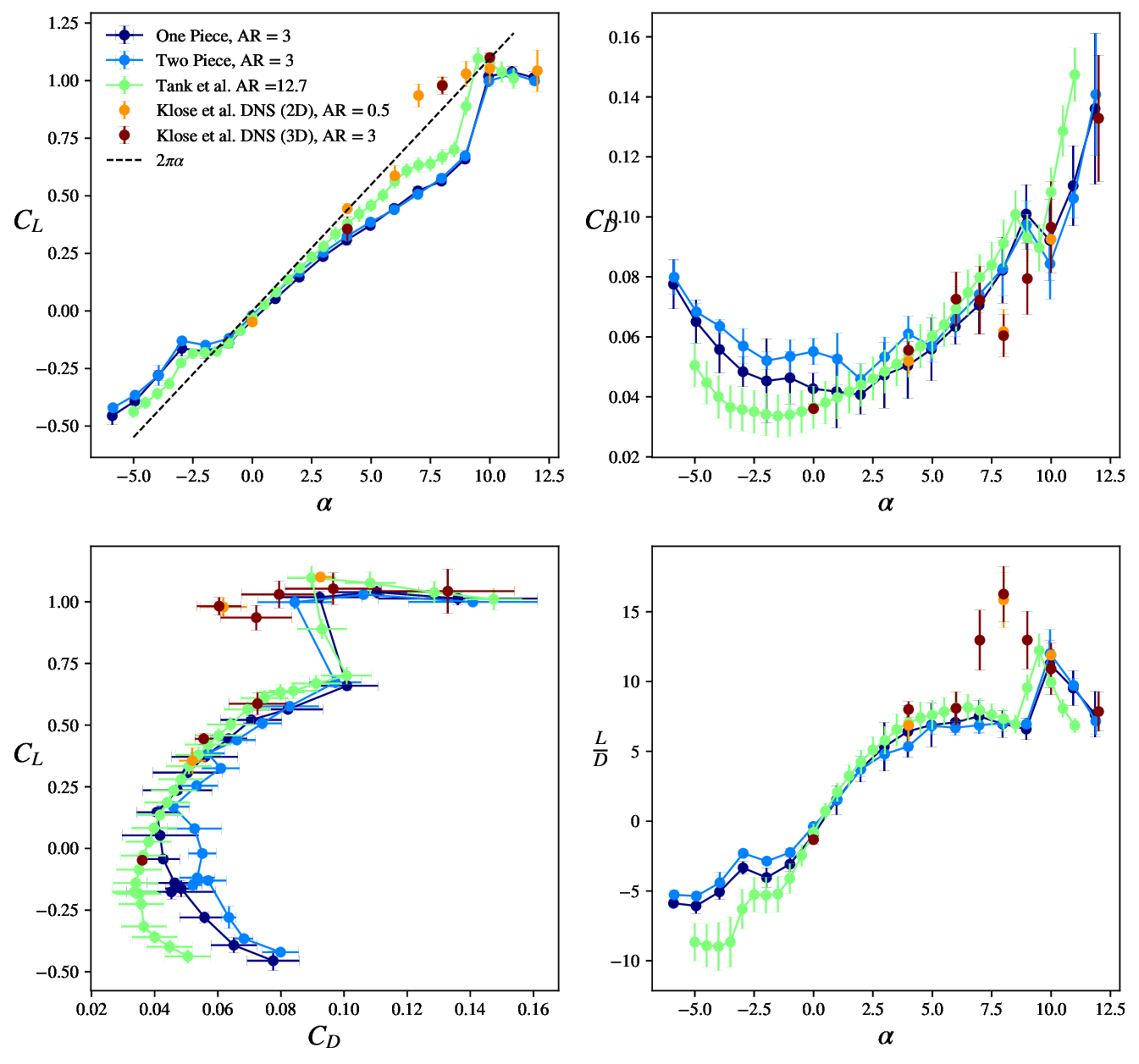}
    \caption{Lift and drag plots at $\mathrm{Re}= 2 \times 10^4$ for different models compared with previous experimental and numerical results. Uncertainties for experiment are the standard deviation of test-test fluctuations of the mean. Uncertainties for simulation are the standard deviation of temporal fluctuations about the mean.}
    \label{fig:Drag Polars Exp vs DNS Re=20k}
\end{figure}

As part of this collaborative work, lift and drag values from three- and two-dimensional DNS were computed for $\RE=2 \times 10^4$. The unforced experiments for each model were compared with previous experimental work \citep{tank:21} and corresponding DNS \citep{klose:25} in figure \ref{fig:Drag Polars Exp vs DNS Re=20k}. Good agreement is shown between experiment and DNS for $\alpha< 7^\circ$ and $\alpha> 10^\circ$. The DNS identifies $\alpha_{\text{crit}} = 7^\circ$ while experiments identify $\alpha_{\text{crit}} = 9^\circ$, leading to major discrepancies for lift and drag over $7^\circ < \alpha < 9^\circ$. The current experiments showed good agreement with previous experiments, demonstrating similar $\alpha_{\text{crit}}$ values. $|C_L|$ from \cite{tank:21} is slightly higher than from the current experiments and may be due to the difference in aspect ratio $AR=3$ (current) vs $AR=12.7$, a difference that \cite{tank:21} also showed to be responsible for a reduction in $\alpha_{\textrm{crit}}$ of $0.5^\circ$. Plotting only the data at $\RE=2 \times 10^4$ allows for a $C_L$ discontinuity at negative $\alpha$ to be seen near $\alpha=-3^\circ$. This is likely due to separation dynamics occurring on the pressure side of the airfoil and does not occur for $\RE\geq 6 \times 10^4$. 

\begin{figure}
    \centering
    \includegraphics[width=1\linewidth]{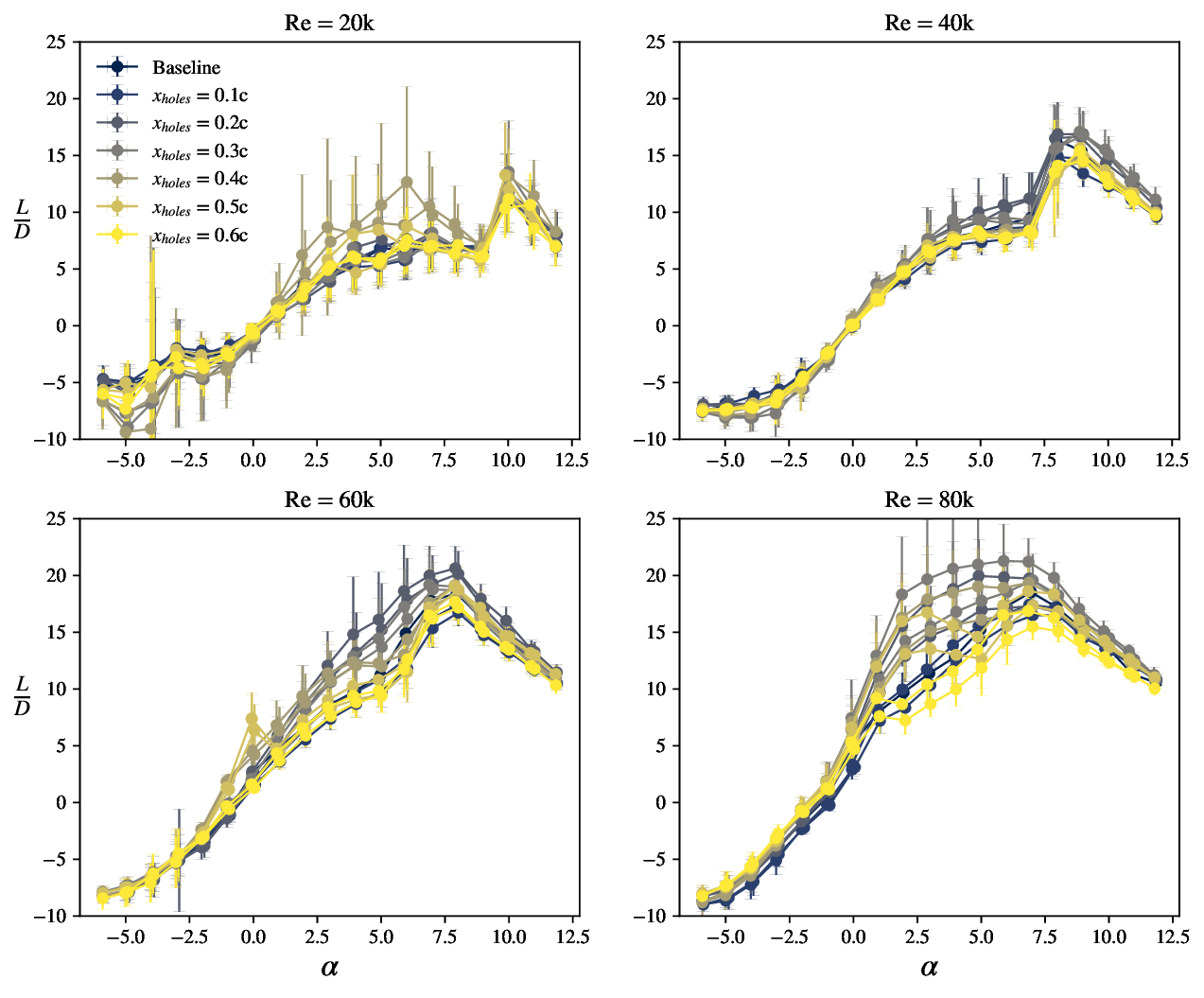}
    \caption{$\Delta L/D$ relative to two--piece model with smooth suction surface (baseline), for the suction surface with spanwise speaker holes (no acoustic excitation) at different chordwise locations ($x_{\text{holes}}$). $\RE=2, 4, 6, 8 \times 10^3$ from top left to bottom right.}
    \label{fig:LoD smooth vs holes Contours}
\end{figure}

To evaluate the aerodynamic effect of introducing orifices for acoustic excitation on the suction surface, a comparison of the baseline two-piece airfoil measurements and the two-piece airfoil with holes at different chordwise locations is shown in figure \ref{fig:LoD smooth vs holes Contours}. For these runs, inactive speakers were placed inside their cavities to minimize Helmholtz resonance effects. At $\RE=2 \times 10^4$ $L/D$ is most affected by $x_{\text{holes}}/c = 0.4$ increasing $L/D$ for $2^\circ \leq \alpha \leq 9^\circ $. For $\RE\geq4 \times 10^4$  locations $x_{\text{holes}}/c = 0.2 - 0.5$ introduce moderate increases in $L/D$ with the effect becoming more pronounced as $\RE$ increases. Locations of $x_{\text{holes}}$ closer to the leading edge have effects which persist over larger ranges of $\alpha$ and increases in $L/D$ experience sharp drop offs at $\RE=6 \;\&\; 8 \times 10^4$ for $x_{\text{holes}}/c \geq 0.3 \;\&\; 0.4$ respectively.

\subsection{Forcing Frequency Identification}
\label{subsec:Forcing Frequency Identification}

\begin{figure}
    \centering
    \includegraphics[width=1\linewidth]{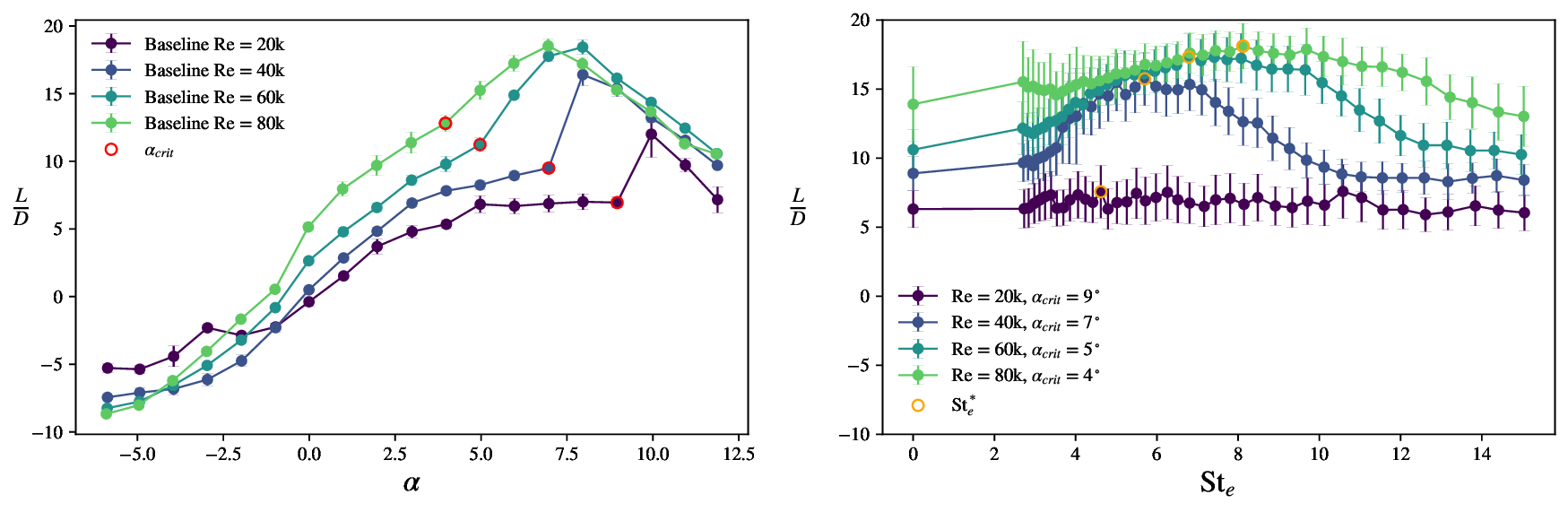}
    \caption{$\alpha$ sweeps of baseline configuration with $\alpha_{\text{crit}}$ identified by red circles for each $\RE$ (left). Frequency sweeps at $\alpha_{\text{crit}}$ with $x_e/c=0.1$ and $\ST_{e}^*$ identified by orange circles for each $\RE$ (right). $\ST_{e}=0$ data point corresponds to unforced measurement at beginning of frequency sweep. }
    \label{fig:Optimal forcing frequencies}
\end{figure}

Sweeps through different forcing frequencies at $\alpha_{\text{crit}}$ for each $\RE$ are shown on the right in figure \ref{fig:Optimal forcing frequencies}. For reference, the baseline $\alpha$ sweeps are shown on the left with $\alpha_{\text{crit}}$ identified. The sweeps were conducted at $\alpha_{\text{crit}}$ where the separated region is the largest, making it the most susceptible to forcing. To isolate the effects of forcing, the forcing location $x_e/c=0.1$ was selected for the initial frequency sweeps since the presence of the speaker holes had little to no effect on the airfoil performance at this location, as shown by figure \ref{fig:LoD smooth vs holes Contours}. The frequency sweeps for $\RE\geq 4 \times 10^4$ generate concave curves of $L/D(\ST)$, while the frequency sweep at $\RE= 2 \times 10^4$ is unaffected by the forcing. The optimal forcing frequency was identified as 
\[\ST_{e}^* = \underset{\text{\ST}_e}{\mathrm{argmax}} \frac{L}{D}.\] 

\begin{figure}
    \centering
    \includegraphics[width=0.5\linewidth]{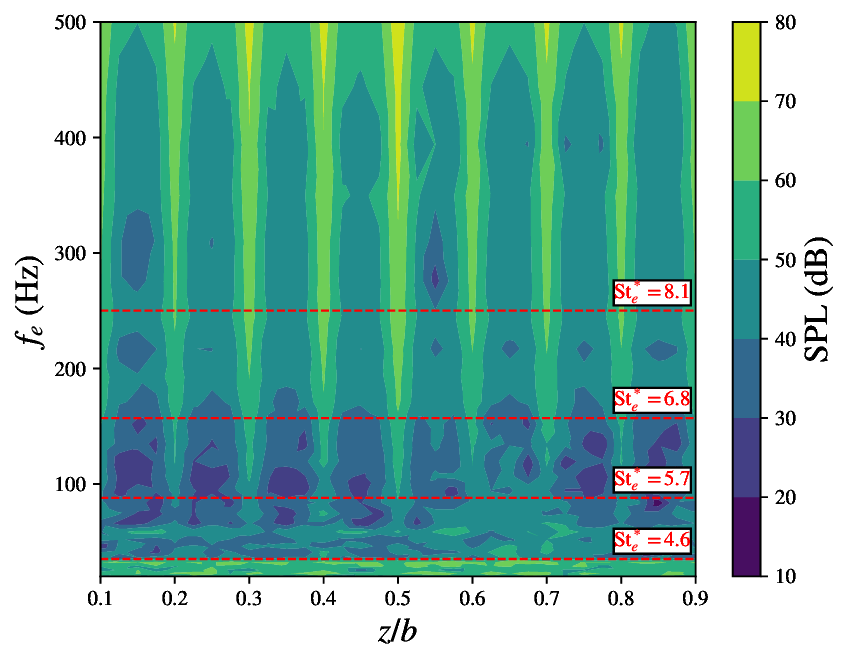}
    \caption{Frequency response of internal excitation with wind tunnel off; measured at a distance of $\leq 0.1$ mm to the suction surface at $x/c=0.3$, with speakers mounted at $x_e/c=0.3$.}
    \label{fig:internal frequency response SPL}
\end{figure}

The UT-P2020 MEMS speakers selected for forcing had a nominally flat amplitude response for frequencies from $f=20 - 700$ Hz according to their data sheet. However, acoustic measurements of the speakers from figure \ref{fig:internal frequency response SPL} showed decreasing SPL with a decrease in $f$. The contour yields ridges of high SPL at the speaker locations every $0.1b$ in the spanwise direction. The highest ridge is at the center location due to the constructive effects of the sound waves. The frequency axis shows diminishing SPL as the frequency decreases, and below 60 Hz the microphone self--noise dominates. $\ST_e^*$ for each $\RE$ is located on the contour and shows that the relative ineffectiveness of forcing for $\RE= 2 \times 10^4$ is likely more related to a reduced forcing amplitude than an intrinsic property of the flow. For the following plots only $\RE= 4-8 \times 10^4$ will be considered.

\begin{figure}
    \centering
    \includegraphics[width=0.5\linewidth]{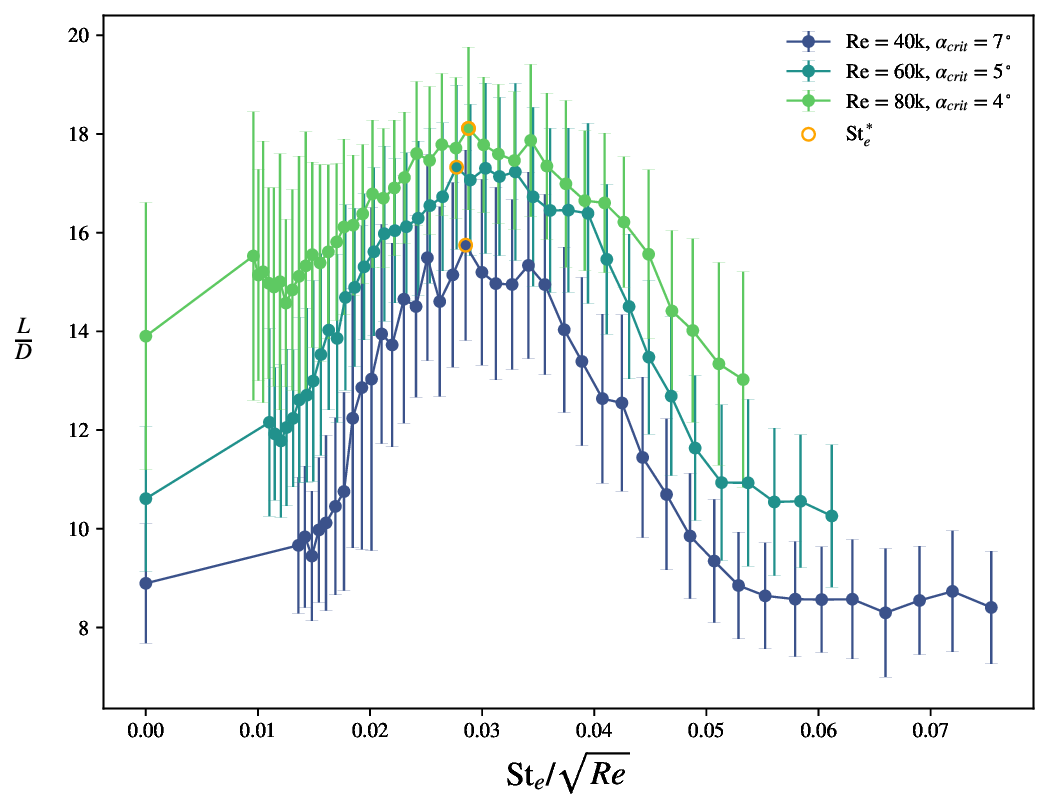}
    \caption{$L/D$ from frequency sweeps with $\STREInline$ scaling \citep{zaman:91}. Uncertainties are the standard deviation of temporal fluctuations about the mean.}
    \label{fig:Optimal forcing frequencies St_Re^0.5}
\end{figure}

Taking the data from the frequency sweeps and applying the scaling proposed by \cite{zaman:91} the optimal rescaled forcing frequencies, $\ST^*_e$, have a constant value $\STeREInline \approx 0.027$. This value agrees with power spectral density measurements taken in the shear layer by \cite{klewicki:25}, which came from the same airfoil at $\alpha = 6^\circ$ and the same Reynolds numbers. This indicates that the forcing is interacting with shear layer vortex shedding, perturbing the system with frequencies near unstable shear layer modes. The proximity of the forcing to the most unstable shear layer modes will dictate temporal and spatial growth rates changing the size and shape of any resulting LSBs. Previous studies \citep{watmuff:99, yarusevych:09, diwan:09, marxen:2012, dellacasagrande:24} have pointed to Kelvin-Helmholtz type instabilities as the primary instability. Physical arguments can be made to relate the $\STeREInline=0.02-0.03$ values for optimal forcing cited by \cite{zaman:91} to Kelvin-Helmholtz instabilities occurring in the laminar separation shear layer. If the most amplified  K-H frequency of $f_{KH} = 0.032\bar{U}/\theta_{SL}$ from \cite{ho:84} is used as the excitation frequency, where $\bar{U} = (U_1+U_2)/2$ is the mean velocity of the shear layer, and $\theta_{SL}$ is the momentum thickness of the shear layer, it is possible to approximate the values from experiment. Assuming $U_2\approx0$, $U_1 = U$, using the flat plate Blasius boundary layer thickness estimate $\theta_{SL}\approx\theta_{FP} = 0.664x_s/\sqrtRExsInline$ for the shear layer momentum thickness, and defining the separation point as some fraction of the chord $x_s = c/n$, where $1\leq n$, then

\begin{align*}
    f_{KH} &= 
    0.032\frac{\bar{U}}{\theta_{SL}} = 0.032\frac{U}{2}\frac{\sqrtRExsDisplay}{(0.664)x_s} = 
    0.032\frac{U}{2}\frac{\sqrtREDisplay\sqrtDisplay{n}}{(0.664)c}, 
    \\
     \STeREDisplay &= 
    \frac{f_ec}{U\sqrtREDisplay} = 
    \frac{f_{KH}c}{U\sqrtREDisplay} = 
    \frac{0.032\sqrtDisplay{n}}{2(0.664)}=
    0.024\sqrtDisplay{n}.
\end{align*}

\begin{figure}
    \centering
    \includegraphics[width=0.75\linewidth]{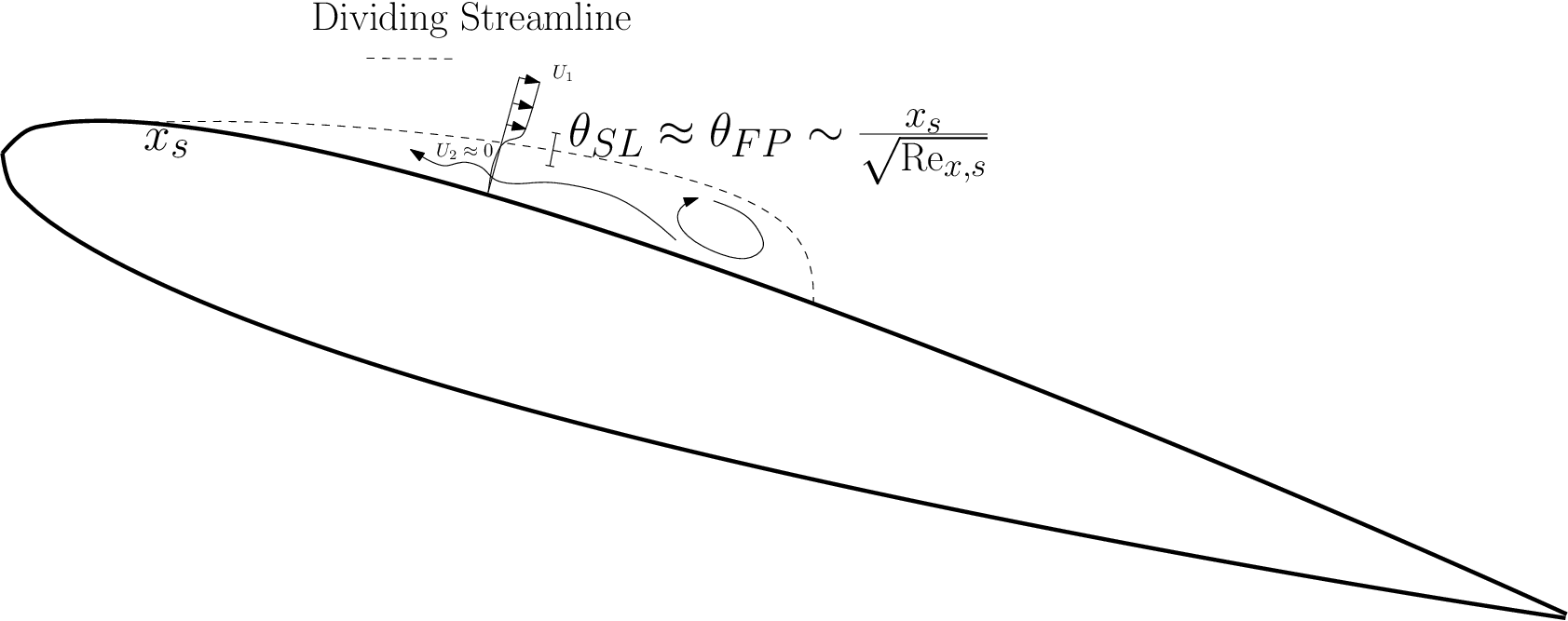}
    \caption{Schematic of laminar separation bubble with shear layer profile.}
    \label{fig:NACA0012_St_Re_Scaling}
\end{figure}

\begin{figure}
    \centering
    \includegraphics[width=1\linewidth]{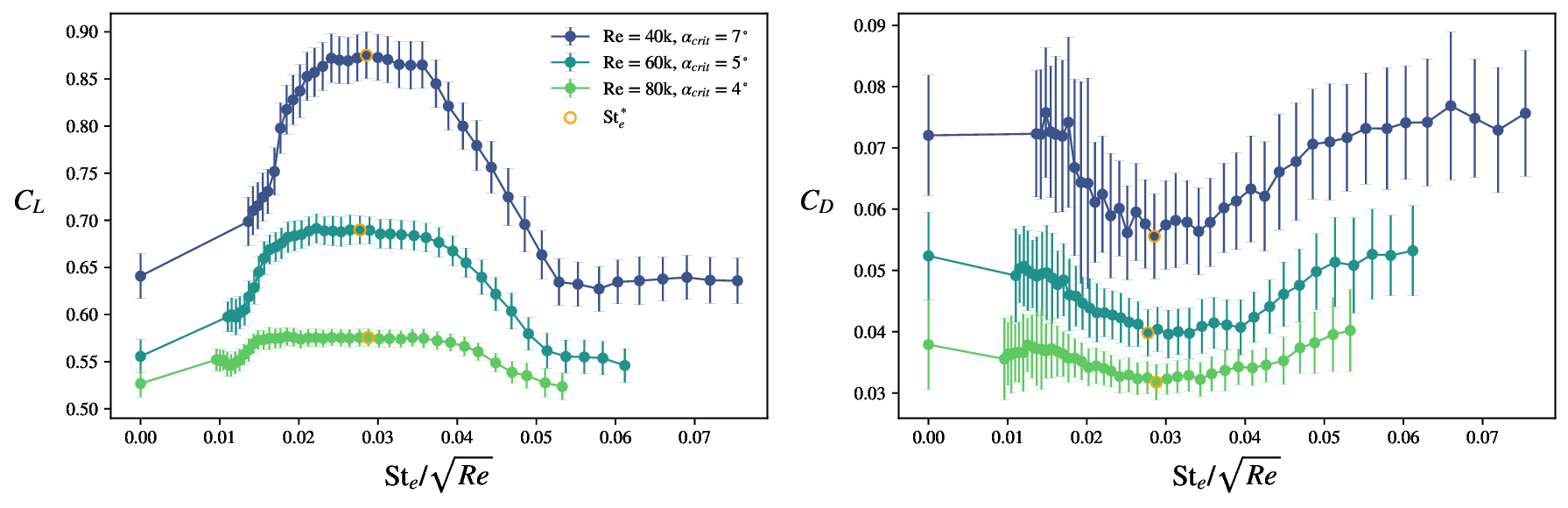}
    \caption{$C_L$ and $C_D$ from frequency sweeps with $\mathrm{St}/\sqrt{\mathrm{Re}}$ scaling \citep{zaman:91}. Uncertainties are the standard deviation of temporal fluctuations about the mean.}
    \label{fig:C_L C_D optimal forcing frequencies St_Re^0.5}
\end{figure}

A naive assumption of $n=1$, where the characteristic length ($x_s$) of the separated shear layer momentum thickness ($\theta_{SL}$) equals the chord length, yields $\STREInline= 0.024$, similar to the value of 0.027 from this study and 0.025 from \cite{zaman:91}. A more empirically based approximation from table \ref{tab:Separation Points} of $n=2$--$8$, where separation is between eighth-chord and mid-chord gives values of $\STeREInline = 0.034-0.067$. The lower end of this range agrees with data from \cite{michelis:18}, \cite{yang:14}, and  \cite{yarusevych:07}. The upper end of this estimate is likely not realized as $\ST_e$ saturates at the K-H frequency when separation takes place at smaller fractions of the chord (higher $\alpha$), causing the shear layer to diverge more quickly from the suction surface. \cite{zaman:91} provide evidence of this with $\STstareREInline = 0.22 - 0.28$ over $\alpha = 0^\circ-6.5^\circ$ for the sharp--nosed LRN airfoil. Overall, this analysis is similar that of \cite{watmuff:99} which compares experimental vortex shedding frequency of a flat plate LSB to canonical linear stability analysis by \cite{monkewitz:82} and two-dimensional vortex shedding work by \cite{pauley:90}. Although the K-H instability is an inviscid instability, for the case of wall-bounded laminar separation, the initial momentum thickness of the shear layer is set by the laminar boundary layer. A cartoon is shown for an airfoil in figure \ref{fig:NACA0012_St_Re_Scaling} and the scaling leads to the collapse of optimal forcing frequencies ($\ST_e^*$) for this study. Most of the studies discussed here fall into the values $n=1-2$ indicating that a length scale between mid-chord and the full chord length will result in an estimate close to whats been observed: $\STstareREInline=0.24 - 0.34$.

Further interrogation of the lift and drag in figure \ref{fig:C_L C_D optimal forcing frequencies St_Re^0.5} shows the lift exhibits a plateau at the center of which the optimal frequency occurs, while drag exhibits convex curves driving the identification of global optimum values. 

\subsection{Forcing Location Trends} 
\label{subsec:Forcing Location Trends}

\begin{figure}
    \centering
    \includegraphics[width=1\linewidth]{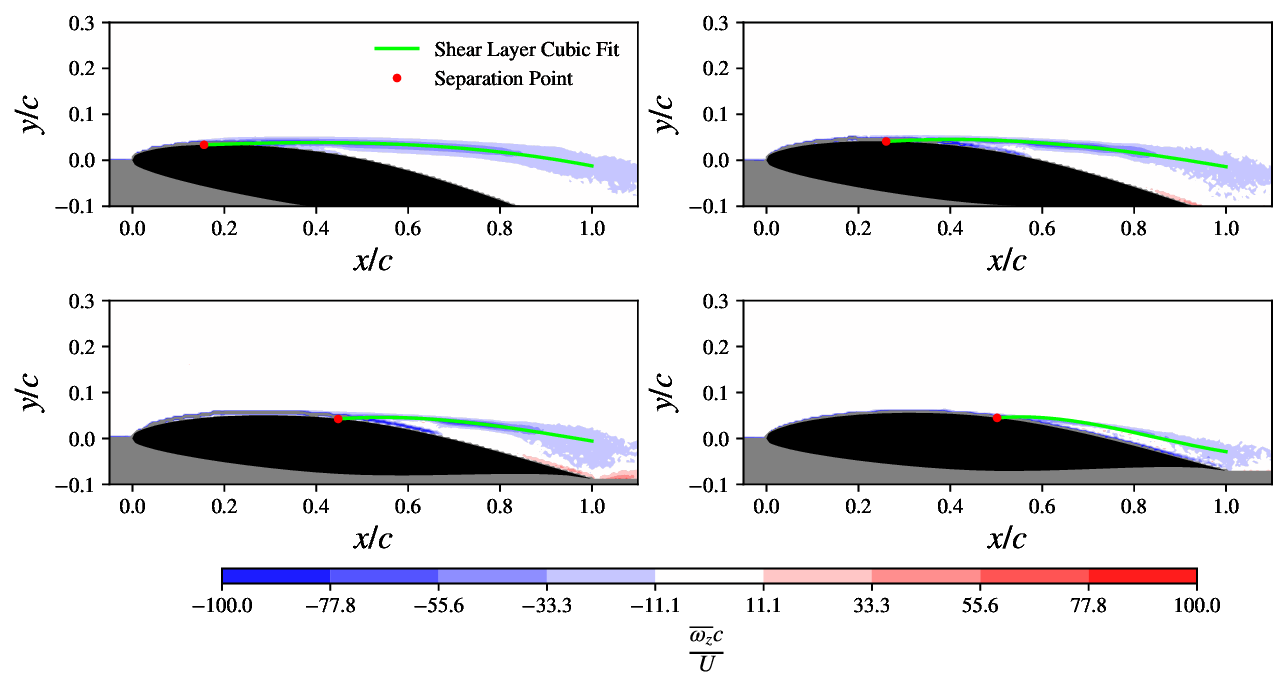}
    \caption{Estimates of separation point. Contours of $\bar{\omega}c/U$ with cubic polynomial fit to shear layer and extrapolated to airfoil surface. ($\RE=2, 4, 6, 8 \times 10^3$ from top left to bottom right.)} 
    \label{fig:Separation Point Estimates}
\end{figure}

Separation point estimates are shown in figure \ref{fig:Separation Point Estimates} and are plotted in the subsequent contours of $x_e$ vs. $\ST_e$. The figure shows that the higher $\alpha_\textrm{crit}$, at lower $\RE$, corresponds to earlier separation and larger separated regions.

\begin{table}
    \centering
    \begin{tabular}{c|ccccccc}
        $\alpha$ & $4^\circ$  & $5^\circ$ & $6^\circ$ & $7^\circ$ & $8^\circ$ & $9^\circ$ & $10^\circ$\\
        \hline
        \underline{Experiment at $\alpha_{\text{crit}}$}  & & & & & & & \\
        $x_s/c$ & 0.50 & 0.45 & - & 0.26 & - & 0.156 & - \\
        \hline
        \underline{DNS at $\RE=2\times10^4$} & & & & & & & \\
        $x_s/c$ & 0.49 & - & 0.4 & 0.26 & 0.014 & - & 0.012\\
    \end{tabular}
    \caption{Separation point locations from DNS \citep{klose:25}  and experiment. $\alpha_{\text{crit}} = 4^\circ,5^\circ,7^\circ,9^\circ$ for experiment correspond to $\RE=8,6,4,2\times10^4$ respectively.}
    \label{tab:Separation Points}
\end{table}

A comparison of the separation points identified by computations and experiment is shown in table \ref{tab:Separation Points}. For $\alpha \leq 7^\circ$ the separation points show good agreement between DNS and experiment.  For $\alpha > 7^\circ$, the DNS at fixed $\RE=2 \times 10^4$ shows the result when $\alpha$ is significantly greater than $\alpha_\mathrm{crit}$ and the separation point has reached its foremost location.

\begin{figure}
    \centering
    \includegraphics[width=1\linewidth]{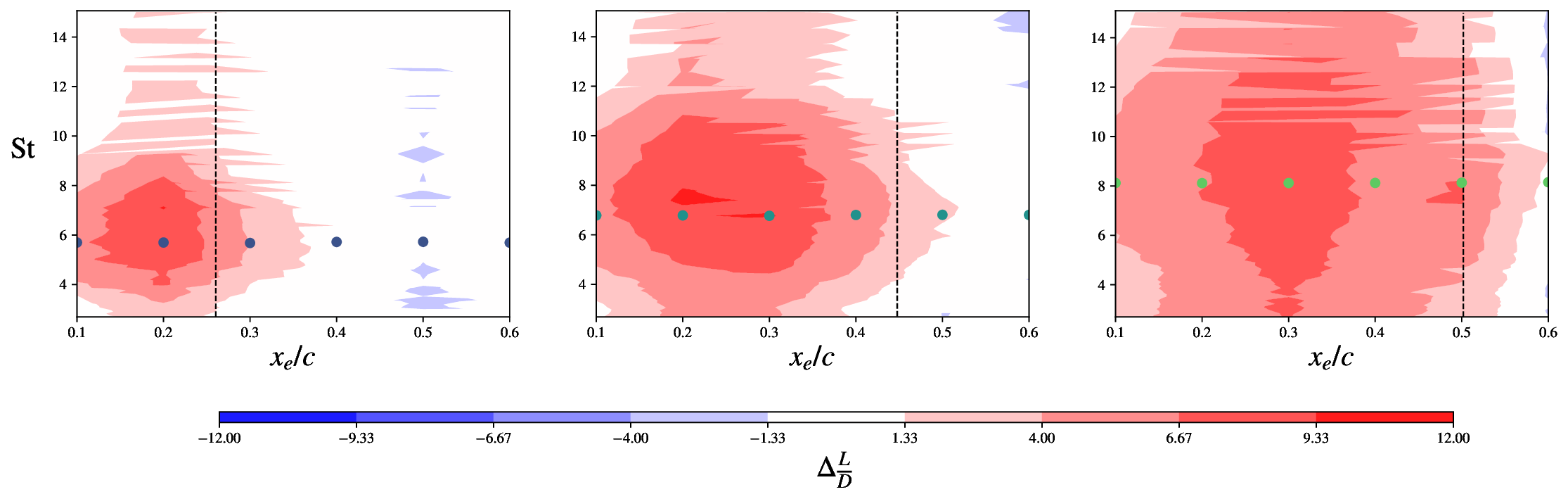}
    \caption{Contours of $\Delta L/D$ relative to smooth baseline from frequency sweeps at different forcing locations $x_e$. Data points at $\mathrm{St}_{e}^*$ for each $x_e$ denoted by colored circles. Black dashed line indicates separation point estimate $x_s$. $\RE=4,6,8 \times 10^4$ at corresponding $\alpha_\text{crit} = 7^\circ, 5^\circ, 4^\circ$ left to right} 
    \label{fig:St Sweep Contours}
\end{figure}

\begin{figure}
    \centering
    \includegraphics[width=1\linewidth]{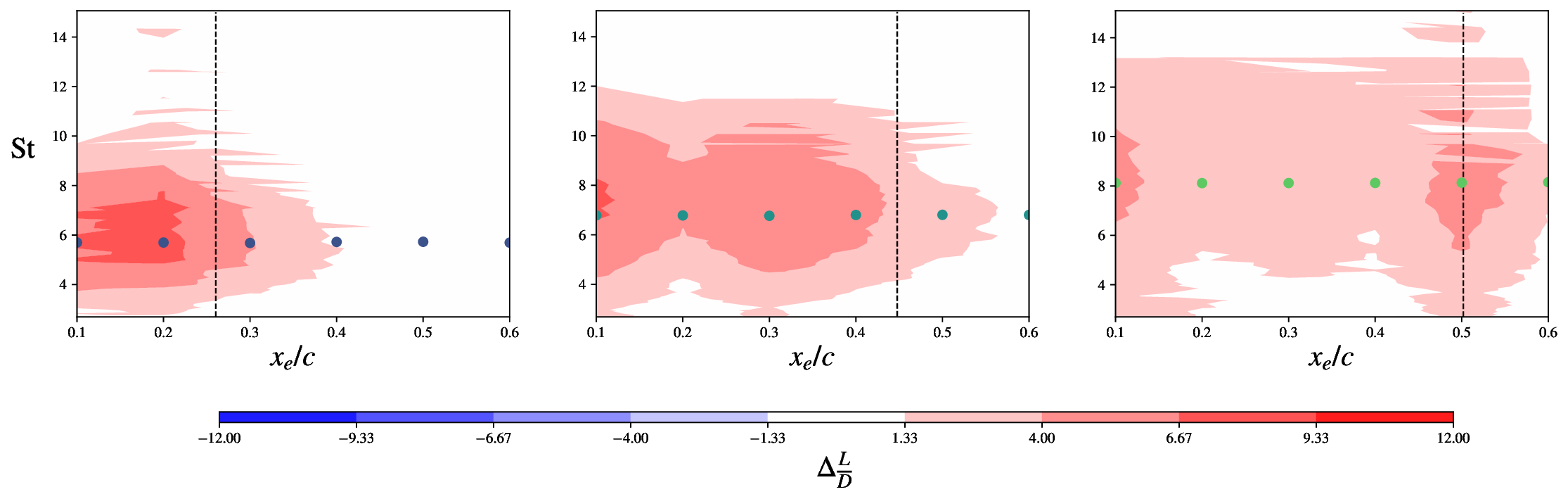}
    \caption{Contours of $\Delta L/D$ relative to suction surface with $x_{\text{holes}}$, from frequency sweeps at different forcing locations $x_e$. Data points at $\ST_{e}^*$ for each $x_e$ denoted by colored circles. Black dashed line indicates separation point estimate $x_s$. $\RE =4,6,8 \times 10^4$ at corresponding $\alpha_\text{crit} = 7^\circ, 5^\circ, 4^\circ$ left to right} 
    \label{fig:St Sweep Contours vs Holes}
\end{figure}

The frequency sweeps from the previous section were carried out at the subsequent chordwise locations $x/c = 0.2-0.6$ and contours comparing the performance with the smooth suction surface (baseline) and the suction surface with speaker holes (no acoustic forcing) are shown in figures \ref{fig:St Sweep Contours} \& \ref{fig:St Sweep Contours vs Holes}. Increases in $L/D$ up to 10 were achieved relative to the baseline and figure \ref{fig:St Sweep Contours vs Holes} shows that acoustic excitation enhanced performance beyond the improved $L/D$ from surface roughness introduced by the speaker holes present on the suction surface. The optimal frequencies ($\ST^*_e$) are indicated in both figures for their respective $\RE$ and show the best performance improvement at each forcing location. This demonstrates that the optimal forcing frequency is independent of forcing location. While some $L/D$ improvement was still observed downstream of the separation line the locations of largest $\Delta L/D$ for both sets of contours all occurred with forcing upstream of separation.

\begin{figure}
    \centering
    \includegraphics[width=1\linewidth]{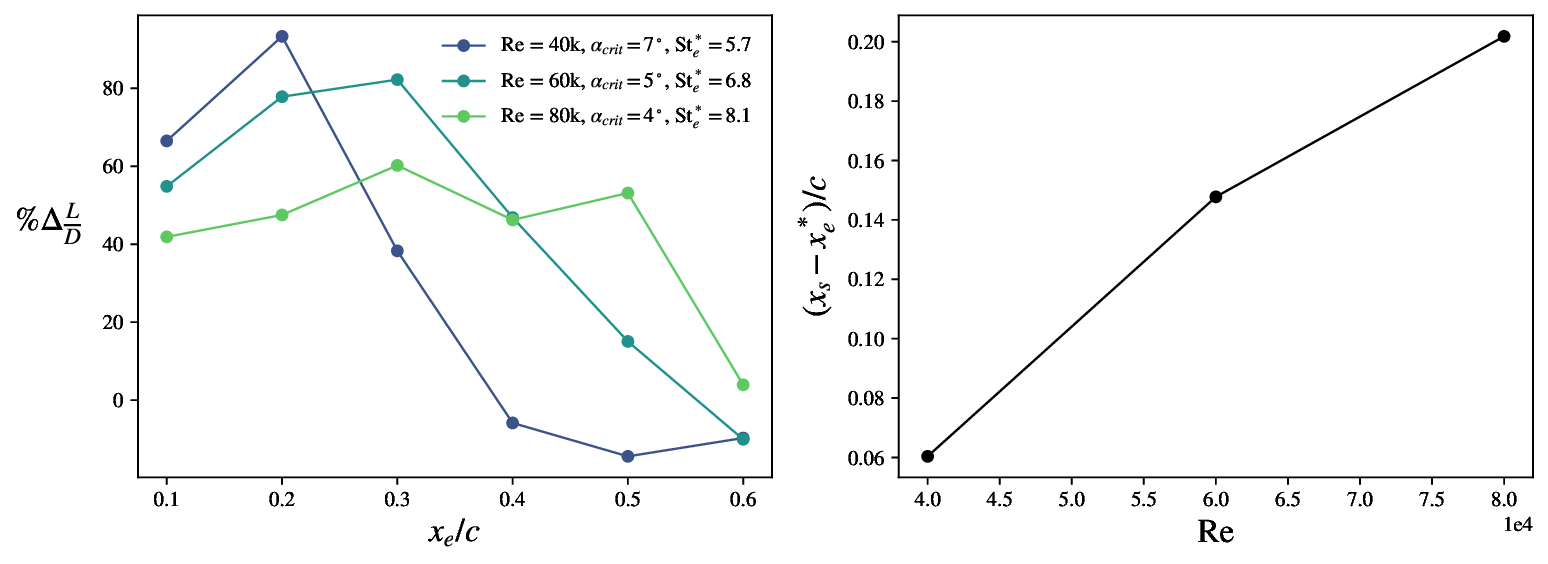}
    \caption{$\%\Delta L/D$ relative to baseline for data points at $\ST_{e}^*$ at each $x_e$ from figure \ref{fig:St Sweep Contours} (left). Difference between separation ($x_s$) location and optimal forcing location $x_{e}^* = \underset{x_e}{\mathrm{argmax}} \frac{L}{D}$ as a function of $\RE$ (right).} 
    \label{fig:St Sweeps max lod vs x_e}
\end{figure}

The individual $\RE$ curves in figure \ref{fig:St Sweeps max lod vs x_e} show an initial increase in the effectiveness of forcing as $x_e$ is moved downstream from $x_e/c = 0.1$, followed by a steep fall-off after reaching their maximum. In the case of $\RE = 4 \times 10^4$, a reduction in $L/D$ occurs for $x_e/c \geq 0.4$. The $x_e$ location with the maximum increase in $L/D$ shifts downstream as $\RE$ increases. The maximum $L/D$ for $\RE = 4 \times 10^4$ occurs at $x_e/c = 0.2$, with near parity between $x_e/c = 0.2$ and $0.3$ at $\RE = 6 \times 10^4$, and a maximum at $x_e/c = 0.3$ for $\RE = 8 \times 10^4$.
These optimal locations $x_{e}^*$ for $\ST_e = \ST_e^*$, where 
\[x_{e}^* = \underset{x_e}{\mathrm{argmax}} \frac{L}{D},\] can be compared with the estimated separation location $x_s$, and the distance upstream of separation at which $x_{e}^*$ occurs is shown on the right in figure \ref{fig:St Sweeps max lod vs x_e}. The location of $x_{e}^*$ moves further upstream of $x_s$ as $x_{e}^*$ increased, perhaps related to the increased amplitude of forcing at higher $\ST_e$, earlier onset of instability, or slower growth rates from unstable modes upstream of separation.

\subsection{Sensitivity of Forcing Frequency and Location to \texorpdfstring{$\alpha$}{alpha}} 
\label{subsec:Sensitivity to alpha}

\begin{figure}
    \centering
    \includegraphics[width=1\linewidth]{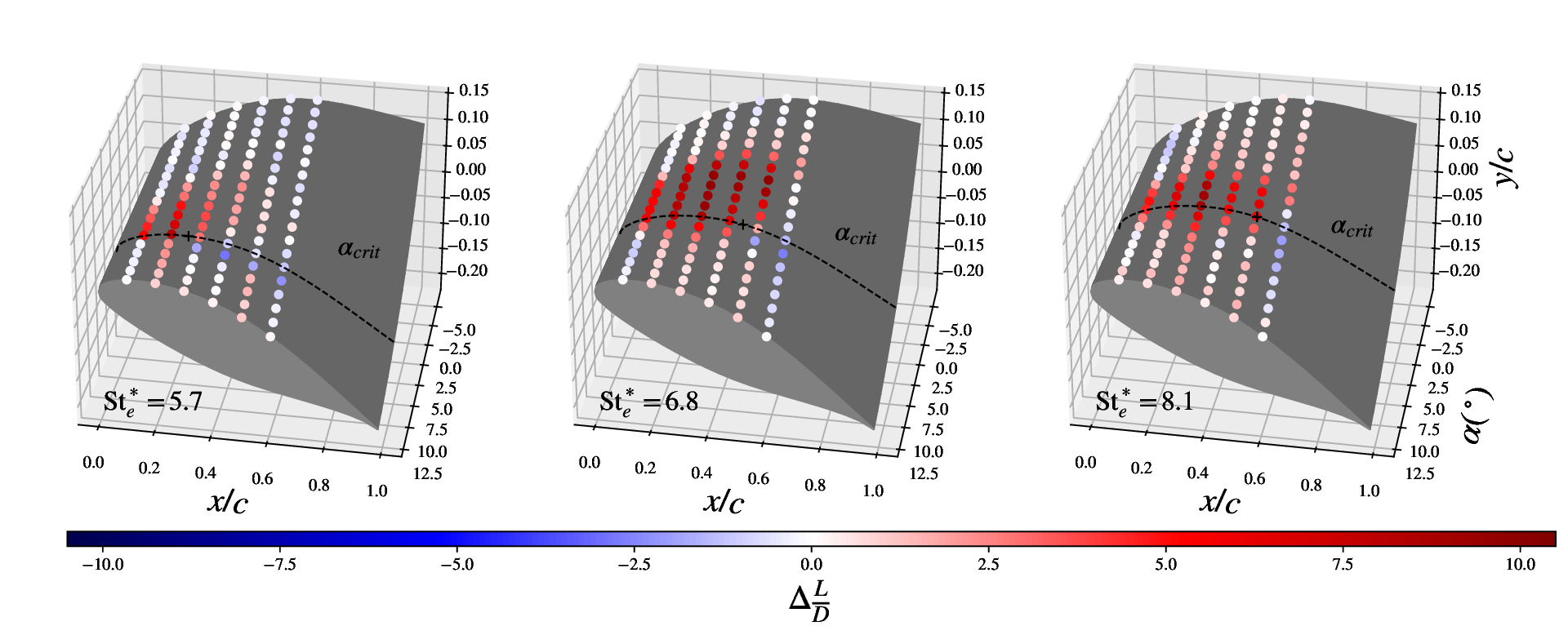}
    \caption{Contours of $\Delta L/D$ relative to baseline from $\alpha$ sweeps at different forcing locations $x_e$. Black dashed lines denote $\alpha_{\text{crit}}$. (+) indicates $x_s$ at $\alpha_{\text{crit}}$.  $\RE = 4,6,8\times 10^4$ left to right.} 
    \label{fig:Alpha Sweep Contours}
\end{figure}

\begin{figure}
    \centering
    \includegraphics[width=1\linewidth]{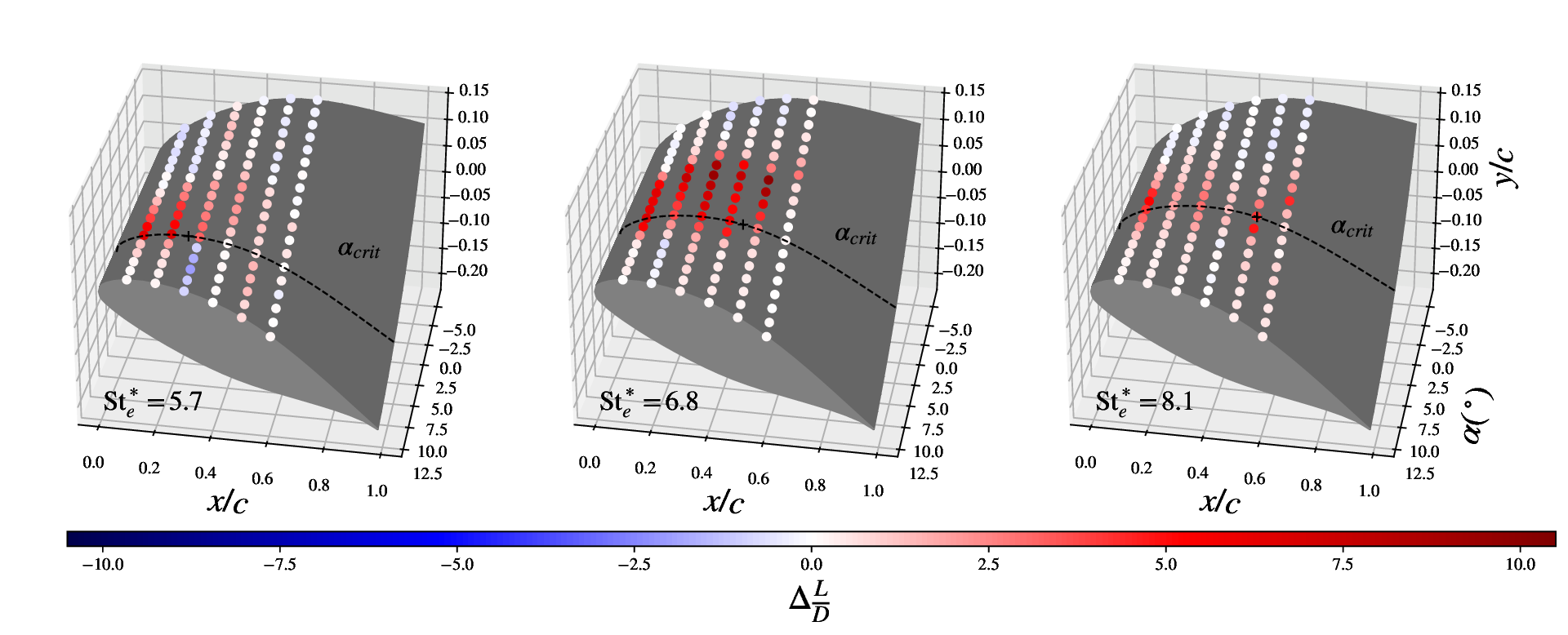}
    \caption{Contours of $\Delta L/D$ relative to suction surface with $x_{\text{holes}}$, from $\alpha$ sweeps at different forcing locations $x_e$. Plotting conventions as previous figure.} 
    \label{fig:Alpha Sweep Contours vs Holes}
\end{figure}

$\alpha$ sweeps for fixed $\ST_e = \ST_e^*$ were conducted and comparisons of $L/D$ to the baseline and suction surface with holes are shown in figures \ref{fig:Alpha Sweep Contours} and \ref{fig:Alpha Sweep Contours vs Holes} so the $\alpha$ parameter appears as a local twist in wing profile. Comparisons with the baseline show improvements over a range of $\alpha$ with peaks in $\Delta L/D$, shown in dark red near $\alpha_\text{crit}$. Improvements in $\Delta L/D$ persist past $\alpha_\text{crit}$, but typically drop off quickly as $\alpha$ increases. The $x_e^*$ locations identified from figure \ref{fig:St Sweeps max lod vs x_e} show improvements in $\Delta L/D$ which persists over the largest range of $\alpha$. Light blue areas of decrease in $\Delta L/D$ occur in all three plots just above $\alpha_\text{crit}$ and downstream of $x_s$. Figure \ref{fig:Alpha Sweep Contours vs Holes} has thinner $\alpha$ bands of $L/D$ improvement compared with those in figure \ref{fig:Alpha Sweep Contours}. The contour shapes are similar to those of figure \ref{fig:Alpha Sweep Contours} with the exception of $\RE=8\times 10^4$, which changes from a contour with a maximum improvement near $x_e/c = 0.3$ at $\alpha_\text{crit}$ to two peaks at $x_e/c = 0.1$ and $0.5$ also at $\alpha_\text{crit}$. These changes in contour shapes can be attributed to the surface impact of holes causing some improvement in $L/D$, but the acoustic forcing significantly increases $L/D$ beyond the improvements from increased surface roughness alone.

\begin{figure}
    \centering
    \includegraphics[width=1\linewidth]{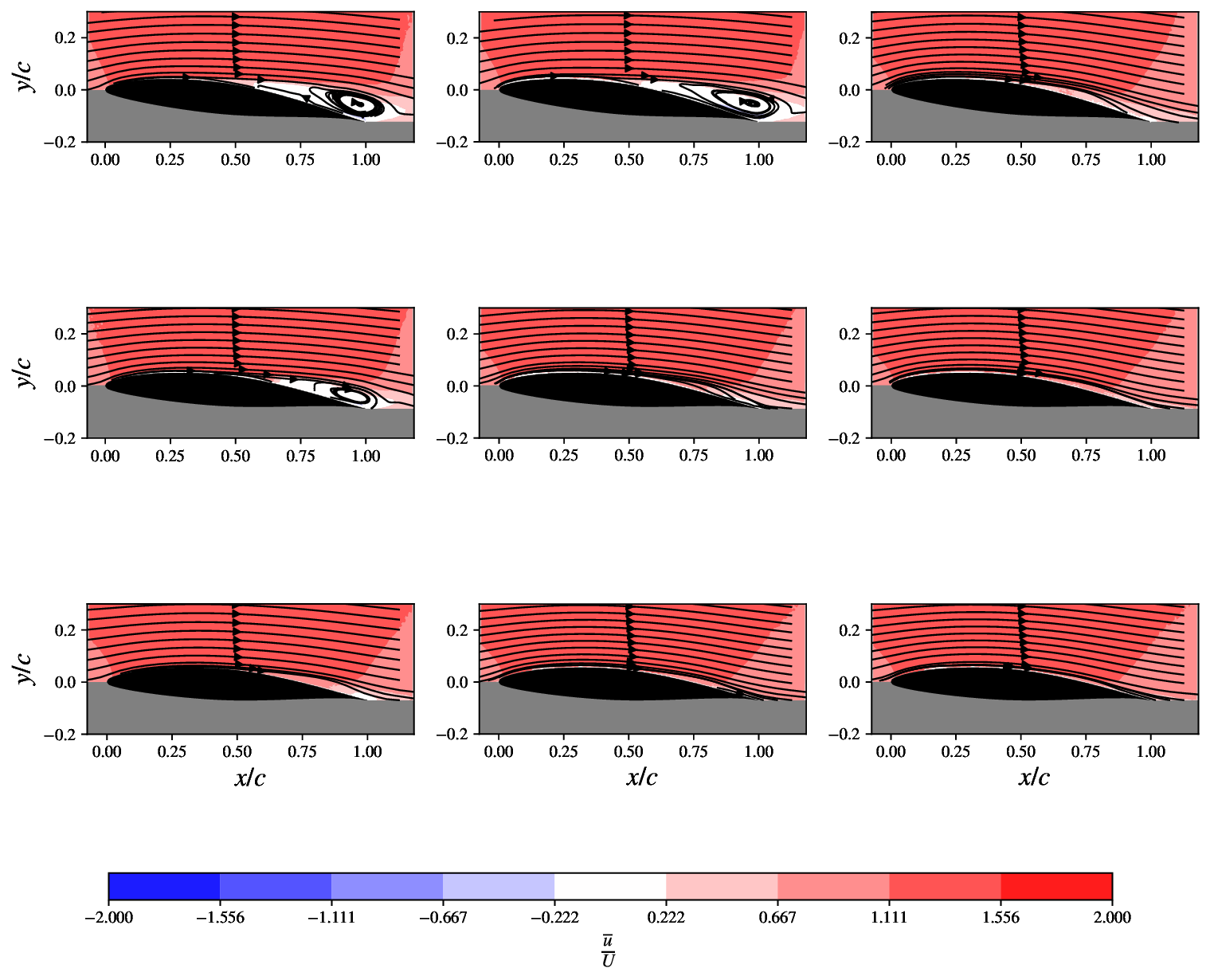}
    \caption{Ensemble averaged PIV $u$ velocity fields with streamlines for unforced one--piece model, unforced two-piece model with speaker holes at $x_{\text{holes}}/c=0.3$, and forced two-piece model with speaker holes at $x_e/c=0.3$ (Left to Right) at $\alpha_{\text{crit}}$ and $\RE = 4,6,8 \times 10^4$ (Top to Bottom).}
    \label{fig:forced vs unforced time averaged u Re=40-80k}
\end{figure}

Changes in ensemble averaged flow fields due to the surface impact of holes and forcing at $x/c=0.3$ can be observed in figure \ref{fig:forced vs unforced time averaged u Re=40-80k}. The unforced flow fields with the smooth suction surface, in the left column of the figure, show separated regions originating near mid-chord and extending to the trailing edge. As $\RE$ increases and consequently $\alpha_\text{crit}$ decreases from top to bottom, the separation region thins and moves downstream. Introducing the speaker holes, shown by the contours in the middle column, has little effect on the $\RE=4\times 10^4$ case (top), but collapses the separated regions for $\RE=6 \,\& \,8\times 10^4$ (middle \& bottom). Similar to the switch in flow state post-$\alpha_\text{crit}$, this change in the flow increases streamline curvature over the chord, increasing lift, and shrinks the cross-sectional area seen by the flow, decreasing drag. Finally, forcing at $x_e/c=0.3$ collapses the $\RE=4\times 10^4$ as well and shrinks the separated regions for $\RE=6 \,\& \,8\times 10^4$, relative to the case with holes and no forcing. The shrinking of the LSB leads to the increase $L/D$ shown in figure \ref{fig:Alpha Sweep Contours vs Holes}. These changes in streamline curvature above the chord and streamwise cross-sectional area experienced by the flow explain the plateau and bowl shapes of $C_L$ and $C_D$ shown in figure \ref{fig:C_L C_D optimal forcing frequencies St_Re^0.5}.

\subsection{Forcing at \texorpdfstring{$Re =2\times10^4$}{=20000} Revisited}\label{subsec:39 speakers}

\begin{figure}
    \centering
    \includegraphics[width=1\linewidth]{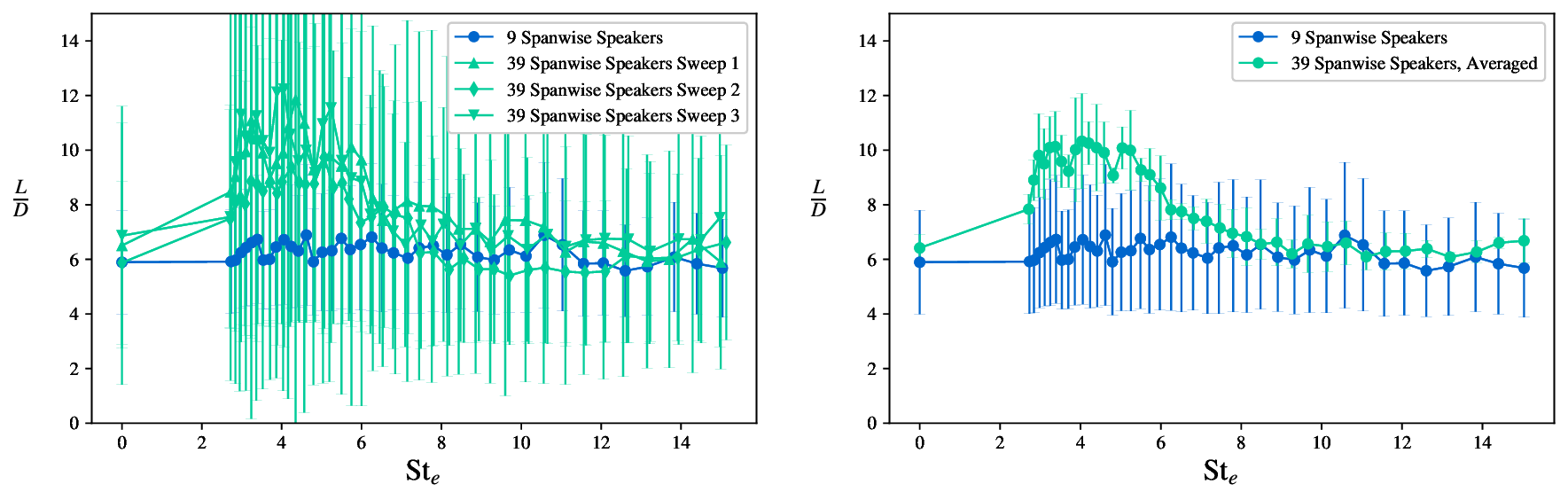}
    \caption{(Left) three individual frequency sweeps at $\alpha_{\text{crit}}$ with $x_e/c=0.1$ and 39 uniformly spaced spanwise speakers, and one frequency sweep with 9. (Right) same as left, but with the three sweeps averaged. The error bars at left show the standard deviations of each individual time series, while on the right, for the 39 speakers, they show run-to-run repeatability (the standard deviation of means).  The $\ST_{e} = 0$ data point corresponds to the unforced baseline at the beginning of each frequency sweep. } 
    \label{fig:39 speakers St sweep}
\end{figure}

Further experiments were conducted to investigate the feasibility of effective forcing at $\RE = 2\times 10^4$.  These experiments were conducted with a higher spanwise density of speaker locations ($z_e/b=0.025-0.975$ by $z_e/b=0.025$) at one chordwise position, $x_e/c= 0.1$. This quadrupling of linear speaker density resulted in 39 speakers, compared with the 9 of the previous wing. Figure \ref{fig:39 speakers St sweep} compares three separate frequency sweeps, each conducted at $\alpha_{\mathrm{crit}}$, as well as their average, with a similar sweep using the 9 speaker wing. Though the error bars are large, in part due to the difficulties in measuring the smaller forces characteristic of this lower Reynolds number, there is a consistent increase in $L/D$, from a baseline $7$ to a maximum of $10$. The plateau shape is in line with similar measurements on the 9 speaker wing at higher Reynolds numbers, as in figure \ref{fig:Optimal forcing frequencies}

The large variance in the individual runs is a measure of the unsteadiness in the flow (in addition to single measurement uncertainty). The increased unsteadiness comes from the vortex formation, convection and shedding over the chord length and from long-period switching between alternate states I and II.  Finally, the plateau of the averaged trace (right) has a maximum at $\ST_e^* = 4.0$, or $\STstareREInline = 0.028$, similar to the $0.027$ value found above.  \par 

\par
For the 9 speaker runs, the original flow response at $\RE = 2\times 10^4$ was weak, likely because the effective forcing amplitudes were low (see figure \ref{fig:internal frequency response SPL}). The stronger response in $L/D$ with roughly four times the number of speakers is due to a combination of 
 increased spanwise coherence of the forcing, together with the higher local excitation amplitudes resulting from greater constructive behavior between discrete stations.

\par
\begin{figure}
    \centering
    \includegraphics[width=1\linewidth]{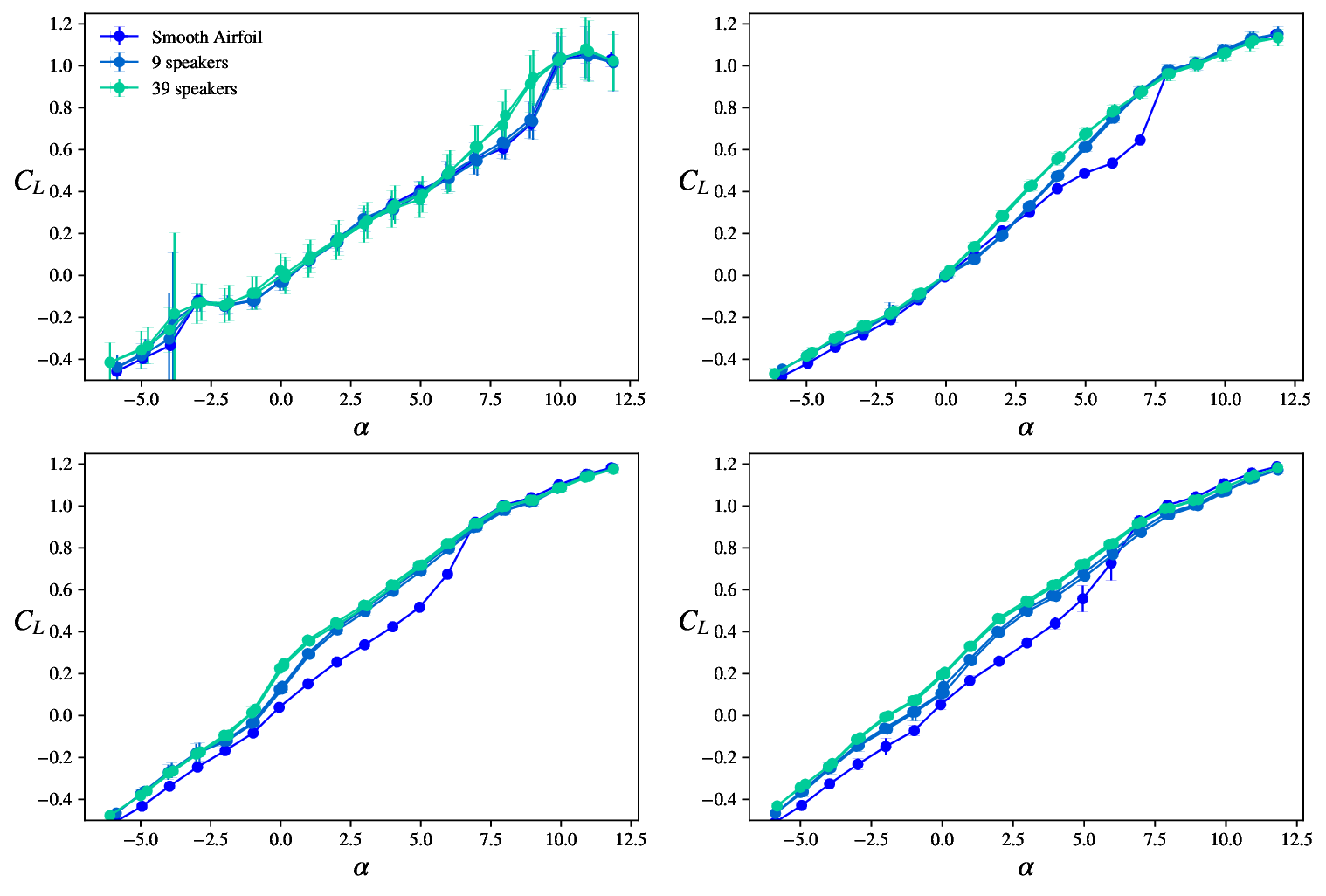}
    \caption{Comparison of $C_L$ vs $\alpha$ for smooth baseline, 9-speaker and 39-speaker wings, forced at $\ST^*$. $\RE=2, 4, 6, 8 \times 10^3$ from top left to bottom right.}
    \label{fig:39 speakers alpha sweeps}
\end{figure}

When forced at $\ST_e^*$, $C_L$ increases over the range $7^\circ \leq \alpha \leq 9^\circ$ in contrast to the lack of measurable increase with 9 speakers.  Similar tests were conducted for the higher $\RE$ (figure \ref{fig:39 speakers alpha sweeps}).  For $\alpha > \alpha_{\mathrm{crit}}$, forcing yields no increase in $C_L$. Similarly, for angles where the 9 speakers already induce a complete switch to SII \citep{yang:14}, increased speaker density gives no additional performance benefit. By contrast, there are lower angles (i.e.  $\alpha\in [0, 1] $ for $\RE = 6\times 10^4$) where the 9 speakers caused a partial switch to SII, while the 39 speakers induce a complete switch. Further, the 39--speaker traces show increased $C_L$ for $\alpha$ where the 9 speakers were ineffective. Interpreting these results in light of the sigmoid-shape $C_L$ vs excitation amplitude curves of \cite{yarusevych:07}, it can be concluded that varying number of speakers affects the flow in a similar manner to that of varying excitation amplitude, and that as $\RE$ or $\alpha$ decreases, the critical excitation amplitude increases.

\section{Discussion and Conclusions}
\label{sec:discussion}

The aerodynamic performance of a wing in the transitional regime ($\RE \in \{10^4, 10^5\}$] has a sensitive dependence on geometry and flow conditions. Small changes in flow speeds and angles of incidence result in the forming and bursting of LSBs with correspondingly abrupt changes in lift and drag. These sensitivities are driven by the easily--destabilized shear layer created by laminar separation, and are persistent over an order of magnitude range in $\RE$. As $\RE$ increases, the boundary layer thins and viscous effects become less influential leading to an increase in $C_L$ as the inviscid limit is approached. Similarly, for increasing $\RE$, there is a decrease in $C_D$ and $\alpha_\text{crit}$, as the flow switches states from laminar separation without time-averaged reattachment to an LSB where the vortices arising from the shear-layer instability act to bring high-speed fluid down once again to the airfoil surface. 

The baseline experiments acquired here and in previous studies \citep{yang:14, michelis:17, tank:21, toppings:24}  highlight the sensitivity of transitional flows to small perturbations, and the purpose of this paper is to show the viability of onboard sinusoidal acoustic forcing on a finite wall-bounded wing, with the objective of identifying favorable forcing frequencies and locations across a range of $\RE$ and $\alpha$.

Experiments demonstrated significant improvement in aerodynamic performance with clear trends for optimal forcing parameters. 
Systematic variation of forcing frequencies ($\ST_{e}$) showed concave curves allowing for identification of optimal forcing frequencies 
\begin{gather}
    \ST_{e}^* = \underset{\ST_e}{\mathrm{argmax}} \frac{L}{D}
\end{gather}
which collapsed across $\RE$ when scaled by $\sqrtREInline$ \citep{zaman:91}. The most amplified frequencies are related to a shear layer instability, similar to a Kelvin-Helmholtz instability \citep{watmuff:99,diwan:09, yarusevych:09}, but seeded by disturbances originating upstream in the boundary layer. If a flat plate momentum thickness, $\theta_{FP}$, is estimated just upstream of the separation location for a lengthscale, then the optimal forcing frequency can be predicted as for an inviscid shear layer instability \citep{ho:84}.

Sweeps through the chordwise forcing location, $x_e$ showed the most effective forcing frequency was always at $\ST_{e}^*$, and that forcing was most effective upstream of separation location, $x_s$. Improvements in $L/D$ diminish quickly downstream of separation, and the upstream distance between optimal forcing, $x_e^*$, and separation, $x_s$, locations increases with $\RE$. 
Further work is necessary to disentangle the roles that forcing amplitude, instability growth rates, and instability onset location play in this last result.

In the lower $\RE$ range,  the shear layer is more robust to surface imperfections, but targeted forcing still causes the separated region on the aft portion of the airfoil to collapse. At $\RE\geq 6 \times 10^4$, surface roughness introduced by the speaker holes themselves was sufficient to collapse--- at least partially---the separated region, passively improving the airfoil performance. Nevertheless, even with LSBs already formed on the airfoil surface, the additional acoustic forcing further improves airfoil performance by reducing the streamwise and wall-normal dimensions of the LSB.

In a uniform flow, far from any boundaries and for a given geometry, the normal and streamwise components of the aerodynamic force depend only on $\RE$, and their ratio, $L/D$, is a convenient measure of performance.  For a fixed airfoil shape, the problem geometry is varied through the angle of attack, $\alpha$, so one expects
\begin{equation}
L/D = f(\RE, \alpha).
\end{equation}
In practice, and especially for $\RE\leq 10^6$, $L/D$ is very sensitive to small departures from these ideal conditions, and ambient turbulence intensity, surface roughness (skin friction, $C_f$), and acoustic environment ($a(A_e,f_e)$), all can modify the flow state, as summarized in eq. \eqref{eq:L/D}.

\begin{align}
    &L/D = f(\RE, \alpha, \underbrace{T, C_f, a(f_e, A_e)\dots}_{\epsilon})  
    \label{eq:L/D}
    \\
    &\epsilon \approx F(f_e, A_e, \vec{x}_e)
    \label{eq:epsilon}
\end{align}

One may think of these departures from ideal flow conditions as perturbations to the base flow set by $\RE$ and $\alpha$, and setup a persistent forcing equivalence (eq. \eqref{eq:epsilon}) which will depend on the excitation frequency ($f_e$), amplitude ($A_e$), and location ($\vec{x_e}$). 

Frequency sweeps of transitional flows by \cite{zaman:91} showed $\Delta C_L$ increases of up to 30\% with a plateau around some preferred frequency. The plateaus collapsed with $\STeREInline$ scaling to $\STeREInline = 0.02 -0.03$ depending on the airfoil. This agrees with the frequency collapse of maximum gains from \cite{rolandi:25} for resolvent analysis across different Re on a NACA0012. The scaling has been extended here to show $(L/D)_{\textrm{max}}$ collapsed to $\STeREInline = 0.027$ for the wall-bounded NACA 65(1)412. Expressing this as a single shear layer constant $C_{SL} \approx 0.025$, (owing to the forcing resonating with shear layer instabilities \citep{diwan:09, rolandi:25}), it is possible to solve for the optimal forcing frequency, $f^*_e$,

\begin{equation}
     C_{SL} = \STstareREDisplay = \frac{f^*_e\sqrt{c\nu}}{U^{3/2}} \Rightarrow f^*_e = U^{3/2}\frac{C_{SL}}{\sqrt{c\nu}}.
     \label{eq:f*}
\end{equation}

From eq. \eqref{eq:f*}, $f^*_e$ for a given airfoil will only be a function of the thermodynamic state and the freestream velocity. For many engineering flows $\nu$ can be assumed to be constant allowing for the optimal forcing frequency to be set by a single measurement of $U$.

Amplitude variation with forcing at the most unstable shear layer frequency for transitional flows has been explored by \cite{yarusevych:07} for an airfoil and \cite{michelis:17} for a flat plate with an imposed non-uniform pressure gradient. The combined measurements demonstrated that increasing the amplitude above some minimum threshold increases $L/D$ for the airfoil and shrinks the separation bubble for the flat plate. Continued increase of the amplitude further increases $L/D$ and shrinks the separation bubble until $L/D$ saturates and the separation bubble is eliminated (at least up to the measurement resolution). This suggests that $(L/D)_{\textrm{max}}$ is reached at an optimal amplitude, $A_e^*$, equal to the saturation amplitude $A_{\textrm{sat}}$.

The most favorable chordwise excitation location for forcing of transitional separated flows has consistently been shown to be at or upstream of the separation location, $x_s$ \citep{hsiao:90, greenblatt:00, yeh:19, yeh:20}. This is in agreement with linear stability analysis of transitional flows which finds the onset of instabilities at or shortly after the onset of the adverse pressure gradient and upstream of separation \citep{diwan:09, michelis:17, yeh:20}. A naive assumption may be the optimal chordwise forcing location is at the onset of instability, as this is where exponential growth of perturbations will occur, but \cite{yeh:20} found with resolvent analysis, that slightly downstream was optimal for a forcing mode magnitude cost function. Likely, determining a chordwise optimal location will depend on the target range of $\alpha$ (which dictates the onset of the adverse pressure gradient and consequently flow instability and separation), and the trade-offs between input energy and performance gain. Here, it has been shown that as long as $x_e \leq x_s$ the bulk of $L/D$ improvement is achieved. From an $L/D$ perspective, as long as the forcing is able to introduce a perturbation at $A_e^*=A_{\textrm{sat}}$ and $\ST_e^*=C_{SL}\sqrtREInline$ downstream of instability onset, then $L/D$ will be maximized. This implies that suboptimal actuator placement can be mitigated by $A_e$ modification as long as $x_e \leq x_s$. \cite{yang:14} provide some evidence for this through consistent improvement in $\Delta\%L/D$ regardless of $x_e/c$ as long as $\STeREInline\approx0.03$.

Literature exploring forcing simultaneously at multiple chordwise locations is limited. \cite{lyon:97} used trips at multiple locations in an effort to reduce LSB drag finding that tripping the flow at a single location always yielded equal or better drag reduction compared with trips at multiple chordwise locations. \cite{yeh:19} examined global forcing vs local forcing at a single chordwise location on a NACA0012 and found that the input-output gain was reduced by an order of magnitude through the restriction of forcing to a 2-D surface at a single chordwise location which spanned the wing surface. Some of this reduction can be attributed to the restriction to a 2-D surface and it should be noted even for the global forcing, the optimal inputs take the form of a continuous chordwise region near the leading edge with a length $<0.05c$.

Optimal spanwise distribution of forcing is still an open question, although recent experimental and numerical studies have begun to explore spanwise structure \citep{michelisThesis:17, yeh:19, yeh:20, toppings:24, rolandi:25, klose:25}.  \cite{michelisThesis:17} demonstrated the development of spanwise modulation of the coherent rollers in an LSB through selected POD modes, and \cite{yeh:19, rolandi:25} showed varying speeds of transition and flow reattachment depending on spanwise modes and commensurate forcing in a resolvent-based analysis. \cite{rolandi:25} performed a biglobal resolvent analysis to investigate the role of spanwise modes on a NACA0012 airfoil at for $\RE\in \{2500, 10000\}$, and distinguished between the influence of shear-layer and wake modes at short and long timescales, respectively. Here, the forcing is continuous in time (as opposed to impulsive \citep{michelis:17}), so a mix of shear-layer and wake modes may be expected, as shown in some detail in \cite{klose:25}. The relative importance of three-dimensional modes can also be affected by compressibility (sometimes used to stabilise numerical solutions) \citep{miotto:22} and by proximity to the wall of the main shear layer modes \citep{liu:23}.

Though the interplay of boundary layer, shear layer and wake instabilities yield a wide range of paths to closed laminar separation bubbles and their stability, the findings reported here are general to wall-bounded transitional flows with applications that extended to both higher and lower chord-based $\RE$ as shown by \cite{yeh:19, coskun:24} and \cite{ribeiro:24}, and boundary layer flows with adverse pressure gradients \citep{michelis:17, celik:23}.

The results described here suggest that significant improvement in $L/D$ across a range of $\RE$, $\alpha$, and different airfoil contours is possible through the introduction of spanwise acoustic excitation at a single chordwise location near the leading edge with most effective frequencies near a normalised value of $\STeREInline = 0.027$.

\bibliographystyle{unsrtnat}
\bibliography{references}  

\end{document}